\documentclass[fleqn,usenatbib]{mnras}

\usepackage[T1]{fontenc}
\usepackage{bbding}
\usepackage[utf8]{inputenc}

\DeclareRobustCommand{\VAN}[3]{#2}
\let\VANthebibliography\thebibliography
\def\thebibliography{\DeclareRobustCommand{\VAN}[3]{##3}\VANthebibliography}

\usepackage{graphicx}	
\usepackage{amsmath}	
\usepackage{amssymb}	

\usepackage{rotating}
\usepackage{pdflscape}
\usepackage{tabulary}                                                           

\usepackage{color}

\usepackage{url}

\usepackage{float}

\usepackage{placeins}

\newcommand{\TESS}{\emph{TESS}}
\newcommand{\Gaia}{\emph{Gaia}}

\title[V994 Herculis: A Triply Eclipsing Sextuple]{V994 Her: A Unique Triply Eclipsing Sextuple Star System}

\author[Zasche et al.]{P.~Zasche$^1$ \thanks{E-mail: zasche@sirrah.troja.mff.cuni.cz},
T.~Borkovits$^{2,3,4,5,6}$, R.~Jayaraman$^7$ , S.~A.~Rappaport$^7$, M.~Bro\v{z}$^{1}$,
\newauthor
D.~Vokrouhlick\'{y}$^{1}$, I.~B.~B\'\i r\'o$^{2,3}$, T.~Heged\"us$^2$, Z.~T.~Kiss$^2$,
R.~Uhla\v{r}$^{8}$,
\newauthor
H.\,M.~Schwengeler$^9$, A.~P\'al$^{4}$, M.~Ma\v{s}ek$^{10}$, S.~B.~Howell$^{11}$,
S.~Dallaporta$^{12}$,
\newauthor
U.~Munari$^{13}$, R.~Gagliano$^{14}$, T.~Jacobs$^{15}$, M.\,H.~Kristiansen$^{16,17}$,
D.~LaCourse$^{18}$,
\newauthor
M.~Omohundro$^{19}$, I.~Terentev$^{20}$, A.~Vanderburg$^{21}$, Z.~Henzl$^{22,23}$,
B.~P.~Powell$^{24}$,
\newauthor
V.~B.~Kostov$^{24,25,26}$\\
$^1$ Charles University, Faculty of Mathematics and Physics, Astronomical Institute, V
Hole\v{s}ovi\v{c}k\'ach 2, Praha 8, \\ \hspace{0.5cm}
180 00, Czech Republic \\
$^2$ Baja Astronomical Observatory of Szeged University, H-6500 Baja, Szegedi \'ut, Kt. 766, Hungary \\
$^3$ ELKH-SZTE Stellar Astrophysics Research Group, H-6500 Baja, Szegedi \'ut, Kt. 766, Hungary \\
$^4$ Konkoly Observatory, Research Centre for Astronomy and Earth Sciences, \\
 \hspace{0.5cm} H-1121 Budapest, Konkoly Thege Mikl\'os \'ut 15-17, Hungary \\
$^5$ ELTE Gothard Astrophysical Observatory, H-9700 Szombathely, Szent Imre h. u. 112, Hungary \\
$^{6}$ MTA-ELTE Exoplanet Research Group, H-9700 Szombathely, Szent Imre h. u. 112, Hungary \\
$^7$ MIT Department of Physics and and MIT Kavli Institute for Astrophysics and Space Research, Cambridge, MA 02139, USA \\
$^8$ Private Observatory, Poho\v{r}\'{\i} 71, 254 01 J\'{\i}lov\'e u Prahy, Czech Republic\\
$^9$ Citizen Scientist, Planet Hunter, Bottmingen, Switzerland \\
$^{10}$ FZU - Institute of Physics of the Czech Academy of Sciences, Na Slovance 1999/2, CZ-182 21, Praha, Czech Republic\\
$^{11}$ NASA Ames Research Center, Moffett Field, CA 94035, USA\\
$^{12}$ ANS Collaboration, c/o Astronomical Observatory, 36012 Asiago (VI), Italy\\
$^{13}$ INAF Astronomical Observatory of Padova, 36012 Asiago (VI), Italy \\
$^{14}$ Amateur Astronomer, Glendale, AZ 85308 \\
$^{15}$ Amateur Astronomer, 12812 SE 69th Place Bellevue, WA 98006, USA \\
$^{16}$ 3DTU Space, National Space Institute, Technical University of Denmark, Elektrovej 327, DK-2800 Lyngby, Denmark \\
$^{17}$ Brorfelde Observatory, Observator Gyldenkernes Vej 7, DK-4340 T\.zll\.zse, Denmark \\
$^{18}$ Amateur Astronomer, 7507 52nd Place NE Marysville, WA 98270, USA \\
$^{19}$ Citizen Scientist, c/o Zooniverse, Department of Physics, University of Oxford, Denys Wilkinson Building, Keble Road,\\ \hspace{0.5cm} Oxford, OX1 3RH, UK \\
$^{20}$ Citizen Scientist, Planet Hunter, Petrozavodsk, Russia \\
$^{21}$ Kavli Institute for Astrophysics and Space Research, Massachusetts Institute of Technology, Cambridge, MA 02139, USA \\
$^{22}$ Hv\v{e}zd\'arna Jaroslava Trnky ve Slan\'em, Nosa\v{c}ick\'a 1713, Slan\'y 1, 274 01, Czech Republic \\
$^{23}$ Variable Star and Exoplanet Section, Czech Astronomical Society, Fri\v{c}ova 298, 251 65 Ond\v{r}ejov, Czech Republic \\
$^{24}$ NASA Goddard Space Flight Center, 8800 Greenbelt Road, Greenbelt, MD 20771, USA \\
$^{25}$ SETI Institute, 189 Bernardo Ave, Suite 200, Mountain View, CA 94043, USA \\
$^{26}$ GSFC Sellers Exoplanet Environments Collaboration
}

 \date{Accepted XXX. Received YYY; in original form ZZZ}
 \pubyear{2023}

\begin{document}
 \label{firstpage}
 \pagerange{\pageref{firstpage}--\pageref{lastpage}}
  \maketitle

\begin{abstract}
We report the discovery with \TESS\ of a third set of eclipses from V994 Herculis  (TIC
424508303), previously only known as a doubly-eclipsing system. The key implication of this
discovery and our analyses is that V994 Her is the second fully-characterized (2+2) + 2 sextuple
system, in which all three binaries eclipse. In this work, we use a combination of ground-based
observations and \TESS\ data to analyze the eclipses of binaries A and B in order to update the
parameters of the inner quadruple's orbit (with a derived period of 1062 $\pm$ 2\,d). The eclipses
of binary C that were detected in the \TESS\ data were also found in older ground-based
observations, as well as in more recently obtained observations. The eclipse timing variations of
all three pairs were studied in order to detect the mutual perturbations of their constituent
stars, as well as those of the inner pairs in the (2+2) core. At the longest periods they arise
from apsidal motion, which may help constraining parameters of the component stars' internal
structure. We also discuss the relative proximity of the periods of binaries A and B to a 3:2 mean
motion resonance. This work represents a step forward in the development of techniques to better
understand and characterize multiple star systems, especially those with multiple eclipsing
components.
\end{abstract}



\section{Introduction}
\label{sec:intro}

\begin{figure*}
\centering
\includegraphics[width=\textwidth]{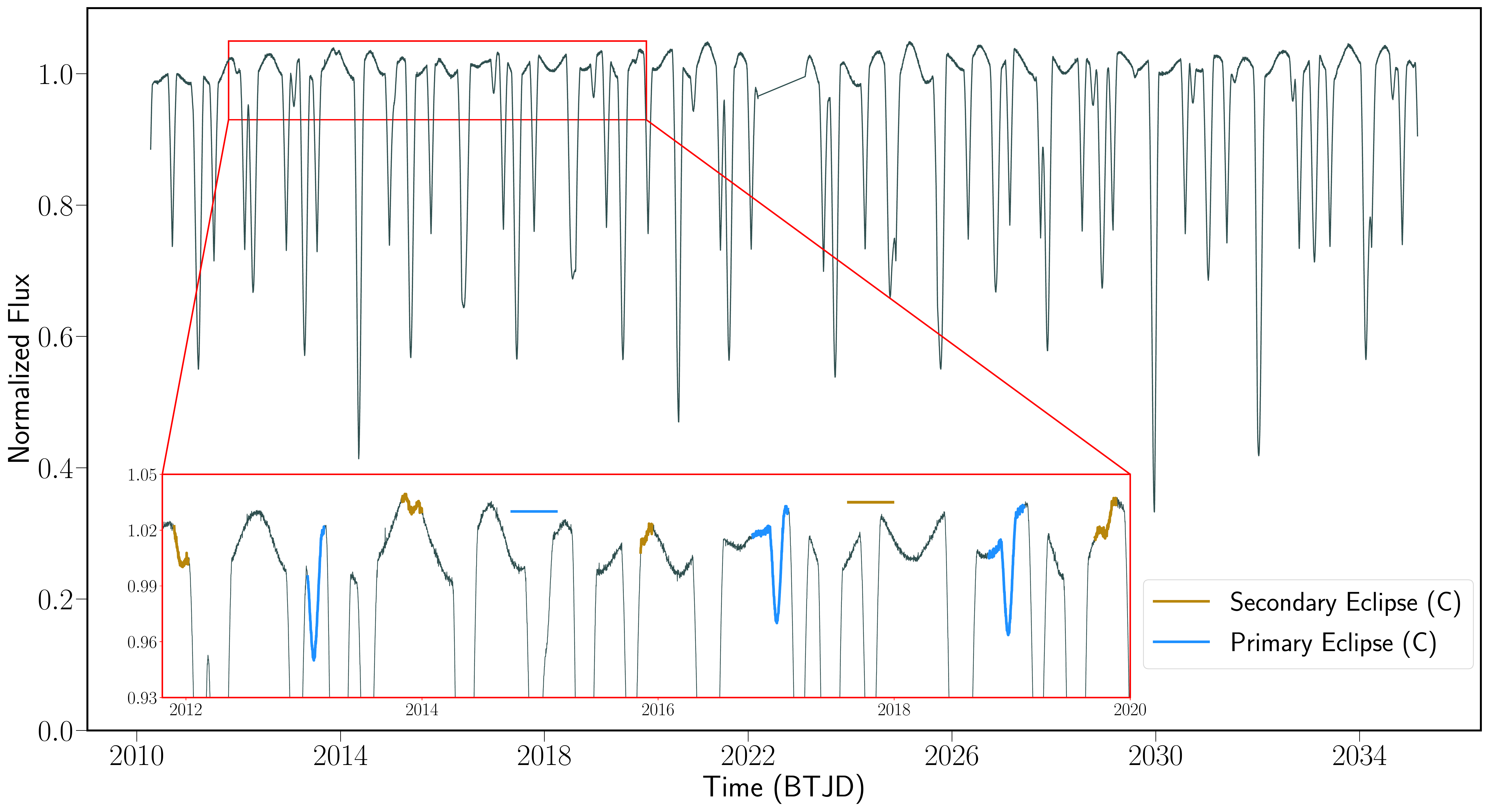}
\caption{The {\em TESS} Sector 26 light curve of TIC\,424508303, aka V994 Her. The x-axis is
plotted in Barycentric TESS Julian Day (BTJD), which corresponds to BJD--2457000.0. The main plot
shows the full 25-day light curve, which includes multiple eclipses from the previously known
eclipsing binaries A and B (\citealt{2008MNRAS.389.1630L,2016A&A...588A.121Z}). It also contains
relatively shallow eclipses from the new binary C, discussed in this work. The inset panel shows a
zoom-in on a roughly 9-d segment of the data.  Three clearly visible primary eclipses of the C
binary are overplotted in blue, while the eclipse lost in a deeper eclipse from the A binary is
indicated with a blue line above its expected location. The (shallow) secondary eclipses of the C
binary are overplotted in gold, with the gold line at BTJD $\sim$ 2017.8 indicating a secondary
eclipse that is lost in one of the deeper eclipses from the ``main'' quadruple.} \label{fig:rawLC}
\end{figure*}   

Multiple star systems consisting of three or more stars are estimated to make up at least 30\% of
binaries, based on a statistical analysis of Kepler data in \citet{2016MNRAS.455.4136B}. However,
only two thousand have been observed in detail, and the number of systems known with
multiplicities higher than 5 is $\lesssim 50$ \citep{Tokovinin2018}. Understanding these
high-multiplicity systems is important, as they can shed new light on many of the open questions
that remain in currently-accepted models of stellar formation and provide insight into the
dynamical interactions of multiple stars \citep{2001ASPC..229...77A}.

V994 Herculis (V994 Her, TIC 424508303) is a bright, well-studied quadruple system, with two known
eclipsing binary components \citep{2008MNRAS.389.1630L}. At the time, this was the first known
doubly-eclipsing quadruple system. The component stars are bright and massive, and the periods of
the two eclipsing binary components have been well-constrained: binary A consists of a B8V and an
A0V star, with a period of 2.083 days; binary B consists of an A2V and an A4V star, with a period
of 1.420 days. These stars are young and occupy a position near the zero-age main sequence (ZAMS)
on the Hertzsprung-Russell (H-R) diagram.

\citet{2013ASSP...31P..39M} initially postulated, based on analyses of photometric data, that this
system could harbor another eclipsing binary. However, because the quality of their data was
rather poor, with relatively low photometric precision, their results were not conclusive enough.
With the advent of high-precision space-based survey missions such as the Transiting Exoplanet
Survey Satellite (\TESS;~\citealt{Ricker2015}), we are able to conclusively confirm the presence
of a third set of eclipses using the \TESS~light curves. Table \ref{tbl:mags} contains basic
information about V994 Her and a nearby visual companion (separated by $\approx 1
^{\prime\prime}$).

\citet{2016A&A...588A.121Z} were the first to accurately constrain the period of the inner
binaries' (A and B) revolution about their common center of mass ($\simeq 1060$ days in their
study). Additionally, they argued that TIC 1685970000, a faint (m$_V$ $\sim 8.8$) neighbor some
1.1$^{\prime\prime}$ away from V994~Her, is also gravitationally bound to the main quadruple,
making it one of the few known quintuple star systems. This putative close companion has been
observed many times since its discovery as a visual double in 1831, and these measurements have
been catalogued in the Washington Double Star Catalog (WDS; \citealt{2001AJ....122.3466M}). The
WDS calls the known quadruple a ``primary star,'' and the fainter companion a ``secondary star.''
However, any physical connection between these two visually-close objects has not yet been
conclusively proven; further follow-up and analyses (such as those in Section \ref{VisualOrbit})
can resolve this question.

\begin{table}
\centering \caption{Archival properties of the V994 Her visual double star}
\begin{tabular}{lcc}
\hline \hline
 Name & V994 Her & TYC 2110-1170-2 \\
  & TIC 424508303 & TIC 1685970000 \\
\hline
RA (J2000, deg) & 276.941222 &  276.941246  \\
Dec (J2000, deg) & 24.697407 &  24.697757\\
\TESS$^a$ & 7.037 $\pm$ 0.017 & 8.3949 $\pm$ 0.6 \\
$B^a$ & 7.136 $\pm$ 0.024 \\
$V^a$ & 6.9599 $\pm$ 0.023 \\
$J^a$ & 6.948 $\pm$ 0.019 \\
$H^a$ & 6.999 $\pm$ 0.0036 \\
$K^a$ & 6.989 $\pm$ 0.023 \\
$W1^b$ & 6.844 $\pm$ 0.07 \\
$W2^b$ & 6.838 $\pm$ 0.02 \\
$W3^b$ & 6.903 $\pm$ 0.018 \\
$W4^b$ & 6.732 $\pm$ 0.067 \\
$G^c$ & 7.0966 & 8.8761 \\
$G_{Bp}^c$ & 6.9898 & 6.9842\\
$G_{Rp}^c$ & 6.9358 & 6.9869\\
Parallax$^c$ (mas) & 3.43639 $\pm$ 0.08394 & 3.48065 $\pm$ 0.09849 \\
PM$^c$ (RA, mas/yr) & 5.5770 $\pm$ 0.0701 & 4.8802 $\pm$ 0.1275 \\
PM$^c$ (Dec, mas/yr) &  11.2675 $\pm$ 0.0765 & 7.1568 $\pm$ 0.0834 \\
\hline
\label{tbl:mags}  
\end{tabular}
\textit{Notes:}  Magnitudes are from (a) TIC-8 catalog \citep{2019AJ....158..138S}. (b) WISE point
source catalog \citep{WISE_catalog}, (c) \Gaia\ DR3; PM stands for proper motion \citep{gaiadr3}.
Some parameters for the visual companion TIC 1685970000 are difficult to come by, as the brighter
primary star is 1.1'' away from it, making measurements difficult.
\end{table}


In this paper, we introduce V994 Her as a bona fide triply eclipsing sextuple star system, which
we identified using \TESS\ data. In addition to the known set of two binaries, the system also
consists of a third binary of period 1.96 days, and we demonstrate that the visual companion
listed in the WDS catalog is likely gravitationally bound to the primary star. In Section
\ref{sec:obs}, we describe all available observational data and how they were prepared and used
for the analysis. Then, Section \ref{sec:modelling} provides detailed modelling of the available
data, while \ref{sec:discussion} discusses the results of our modeling. Finally, in Section
\ref{disc} we discuss the possible architecture of the whole system that we infer from our
findings and comment on the proximity of the inner 2+2 component (binaries A and B) to their
mutual 3:2 mean motion resonance.


\section{Observations of V994 Her}
\label{sec:obs}

\subsection{\TESS\ Observations}
\label{sec:observ}

V994\,Her was observed by \TESS\ during Year 2 in Sector 26 (i.e., June 2020), and during Year 4
in Sectors 40 and 53 (i.e. July 2021 \& June 2022). In Sector 26, this star was observed at
2-minute cadence; this light curve was preprocessed and detrended by the Science Processing
Operations Center (SPOC) pipeline \citep{jenkins_spoc}, which is partially based on that used for
Kepler data. The detrended SPOC light curve from Sector 26 is shown in Figure~\ref{fig:rawLC}. For
the Year 4 observations, however, only the full-frame image (FFI) data (at 10-minute cadence) are
available. These data were processed using the convolution-based differential image analysis
methods of the \texttt{fitsh} package \citep{2012MNRAS.421.1825P}. V994 Her's triply eclipsing
nature was identified both algorithmically and through a visual survey\footnote{This search makes
use of the {\tt LcTools} desktop application \citep{Schmitt2019} to view and study light curves.}
of all stars brighter than 13.5 mag in the \TESS\ FFIs (for more information on the latter
initiative, see \citealt{vsg_article}).

\subsubsection{Three methods for disentangling} \label{sec:disentanglement}

\begin{figure*}
\begin{center}
\includegraphics[width=0.99\textwidth]{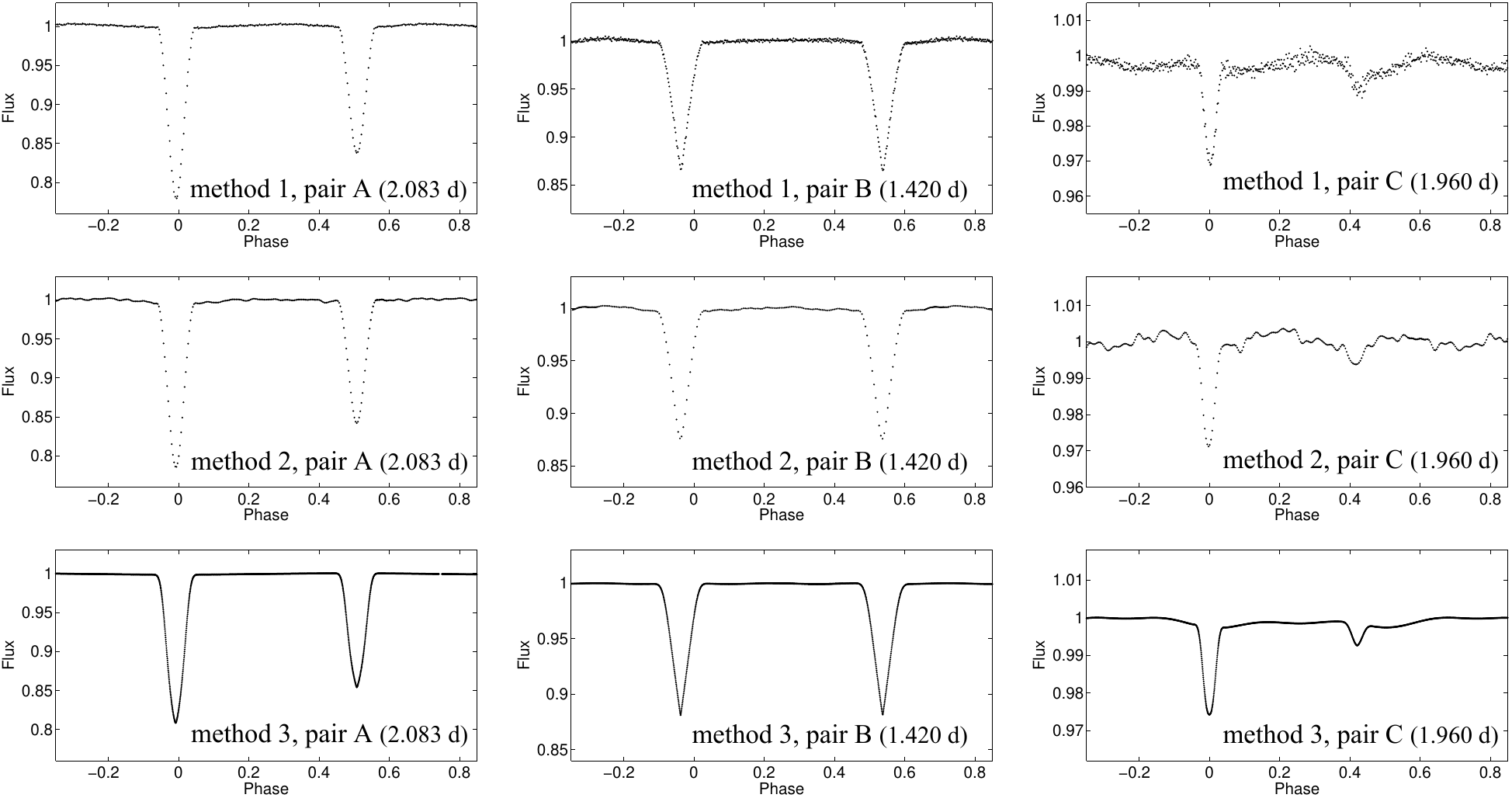}
\caption{The disentangled and folded light curves of all three eclipsing binaries A, B, and C
using different approaches: the time-domain iterative disentanglement (i.e. method 1), the
Fourier-decomposition (method 2), and the iterative phenomenological model methods (method 3),
respectively. For details see the text.} \label{fig:LCs}
\end{center}
\end{figure*}

Using the \TESS\ data, we applied three different methods to disentangle the combined light curve
into the three component eclipsing signals: the time-domain iterative disentanglement method, the
Fourier-decomposition method, and the iterative phenomenological model method. Results for all
three methods are plotted side-by-side in Figure \ref{fig:LCs}.

First, we used the method of time-domain iterative disentanglement. This technique is a powerful
tool for separating the light curves of strongly blended targets, and was described in detail in
Section~3 of \citet{powell_sextuple}, where it was applied for the first time to disentangle the
blended light curves of three eclipsing binaries.

To verify our results and compare the three methods, we used two other methods to disentangle the
light curves. The second one is the Fourier-based iterative method, which was also described in
Section 3.1 of \citet{powell_sextuple}. Such a technique is suitable for these data because all
the signals of interest are strictly periodic over at least one sector of \TESS\ data, wherein
movement on the longer outer orbit can be neglected. The third and final method is based on
iteratively fitting the individual pairs with their respective phenomenological models and then
subtracting these from the overall light curve. After a few (usually two to five) iterative steps,
a shape for the eclipsing light curve of the C pair was clearly obtained. The method itself and
the code used here are described in the Appendix of \cite{2022arXiv220612456P}.

Apart from the three eclipsing signals, the light curve also exhibits an additional pulsation-like
oscillation. Such a variation shows a periodicity of $(P_\mathrm{A}-P_\mathrm{B})$. This extra
feature is apparently not present in the Fourier-disentangled light curve, as well as in the
phenomenologically-disentangled one (see the bottom panels of Figure~\ref{fig:LCs}). This may be
due to the subtraction of this signal as part of the disentanglement process. Unfortunately, we
have not yet been able to come up with a coherent astrophysical explanation of this signal.

We found the first method of time-domain iterative disentanglement as the most suitable for a
subsequent analysis of the individual light curves, which is discussed in Section
\ref{sec:modelling}. This is mainly due to problematic fitting of outside-eclipse parts of the
light curves by the methods 2 and 3.

To derive the precise times of eclipses of each binary, we used the result of the time-domain
iterative disentanglement method. These were calculated for each binary after subtraction of the
light curves of the other two pairs. Eclipse times of each binary, as observed in \TESS, are
presented in Tables \ref{Tab:V994_Her_A_ToM},  \ref{Tab:V994_Her_B_ToM}, and
\ref{Tab:V994_Her_C_ToM}.

\subsection{Ground-based photometric Observations}

\subsubsection{Baja Astronomical Observatory, Hungary (2007)}
\label{sec:Baja2007obs}

V994 Her was observed with the 50-cm f/6 modified Cassegrain Baja Astronomical Robotic Telescope (BART-1), located at the Baja Astronomical Observatory in Hungary, on 40 nights between 18 June 2007 and 9 October 2007. The observations were carried out with a $4096\times4096$ Apogee Alta U16 CCD camera, using a standard Johnson V filter. 

The original goal of this photometric monitoring of V994 Her was to prove and publish for the
first time the previously-unknown doubly eclipsing nature of this system; however,
\citet{2008MNRAS.389.1630L} independently discovered and characterized this system's true nature.
Thus, our team at the time chose to not further analyze the data and simply published the derived
times of minima in \citet{2011IBVS.5979....1B}. However, we make use of this archival photometric
data set in the present work, as it is especially useful for an additional constraining of the
apsidal advance rates of binaries A and B through the complete lightcurve fittings and, also for
checking the constancy (or variability) of the eclipse depths within a one and half decade-long
interval.

\subsubsection{Additional observations}
\label{sec:NewObs}

V994 Her was monitored over several dozens of nights by R.U. at his private observatory in the
Czech Republic, as well as remotely from northern Italy using three different telescopes: a 34-mm
refractor, a 150-mm reflector, and a 200-mm reflector. Some of these observations were obtained
using filtered photometry (usually with $R$ or $I$ filters), while others were carried out without
any filter. Due to the different instrumental setups of these instruments, the comparison stars
were different for each telescope; however, they were always chosen to be adequately close to the
target and of a similar spectral type in order to minimize the effect of differential extinction
during the nights. Additionally, four more nights of data were obtained by the 250-mm
F/(Ph)otometric Robotic Atmospheric Monitor (FRAM) telescope CTA-N, located on the island of La
Palma, Spain \citep{2019ICRC...36..769P}. We also have data from one night of observations by M.M.
at his private observatory in the Czech Republic, using a 200-mm reflector. From the combination
of these datasets, more than 40 new times of eclipses for pair A were derived, and more than 30
for pair B. Several new estimates for pair C were also calculated; however, these are of lower
quality due to the significantly lower photometric amplitude of its variation.

Between 2002 June 10 and 2004 July 14, a total of 1170 measurements in $V$-band and 653 in
$B$-band were collected for V994~Her by S.D. and U.M., using a 28-cm telescope located in Cembra
(Trento, Italy). This telescope was equipped with an Optec SSP-5 photoelectric photometer and
Johnson $B$ and $V$ filters. The comparison and check stars were, respectively, HIP~89975
($V=6.978$~mag, $B-V=-0.095$~mag) and HIP~90637 ($V=5.862$~mag, $B-V=-0.099$~mag). These stars are
nearly identical in $B-V$ color to V994 Her and are located nearby on the sky ($\leq$2$^\circ$
angular separation). From these data, we were also able to derive several times of eclipses for
both the A \& B pairs. Moreover, we were able to derive a rough value for the times of eclipse for
pair C, which allowed us to significantly improve our estimate of its orbital period due to the
increased time coverage.

\begin{figure}
\begin{center}
\includegraphics[width=0.49\textwidth]{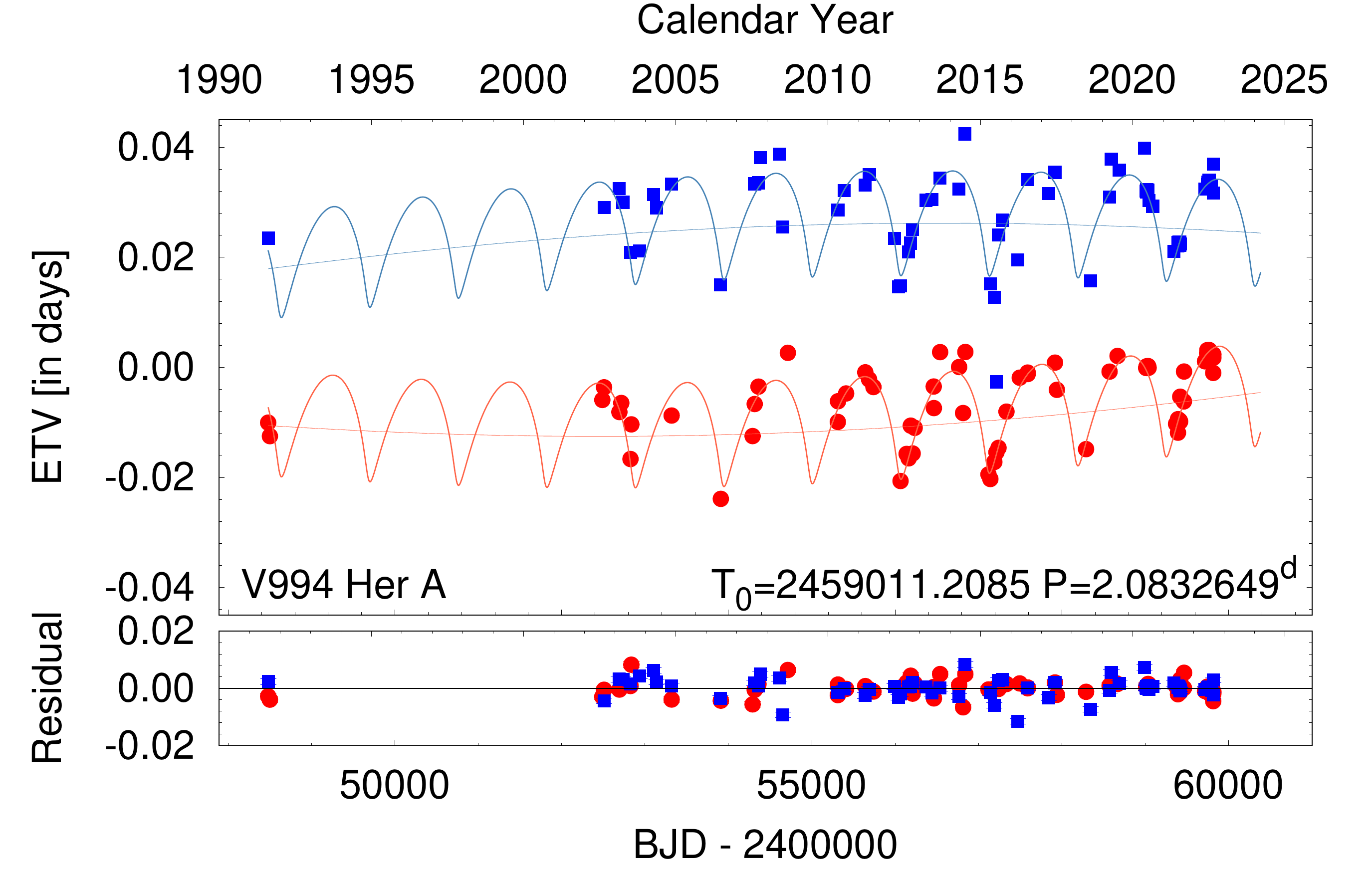} 
\includegraphics[width=0.49\textwidth]{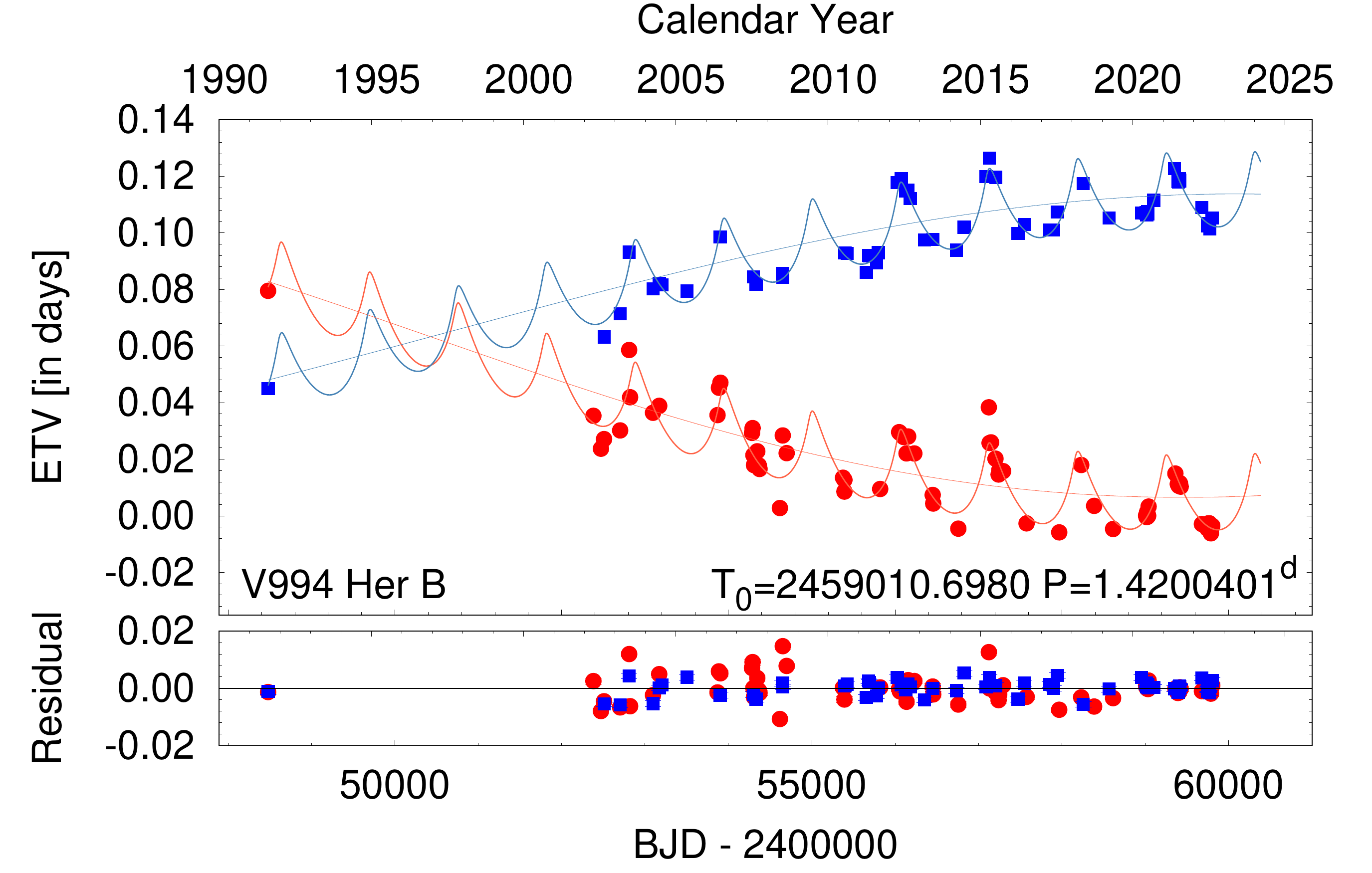} 
\includegraphics[width=0.49\textwidth]{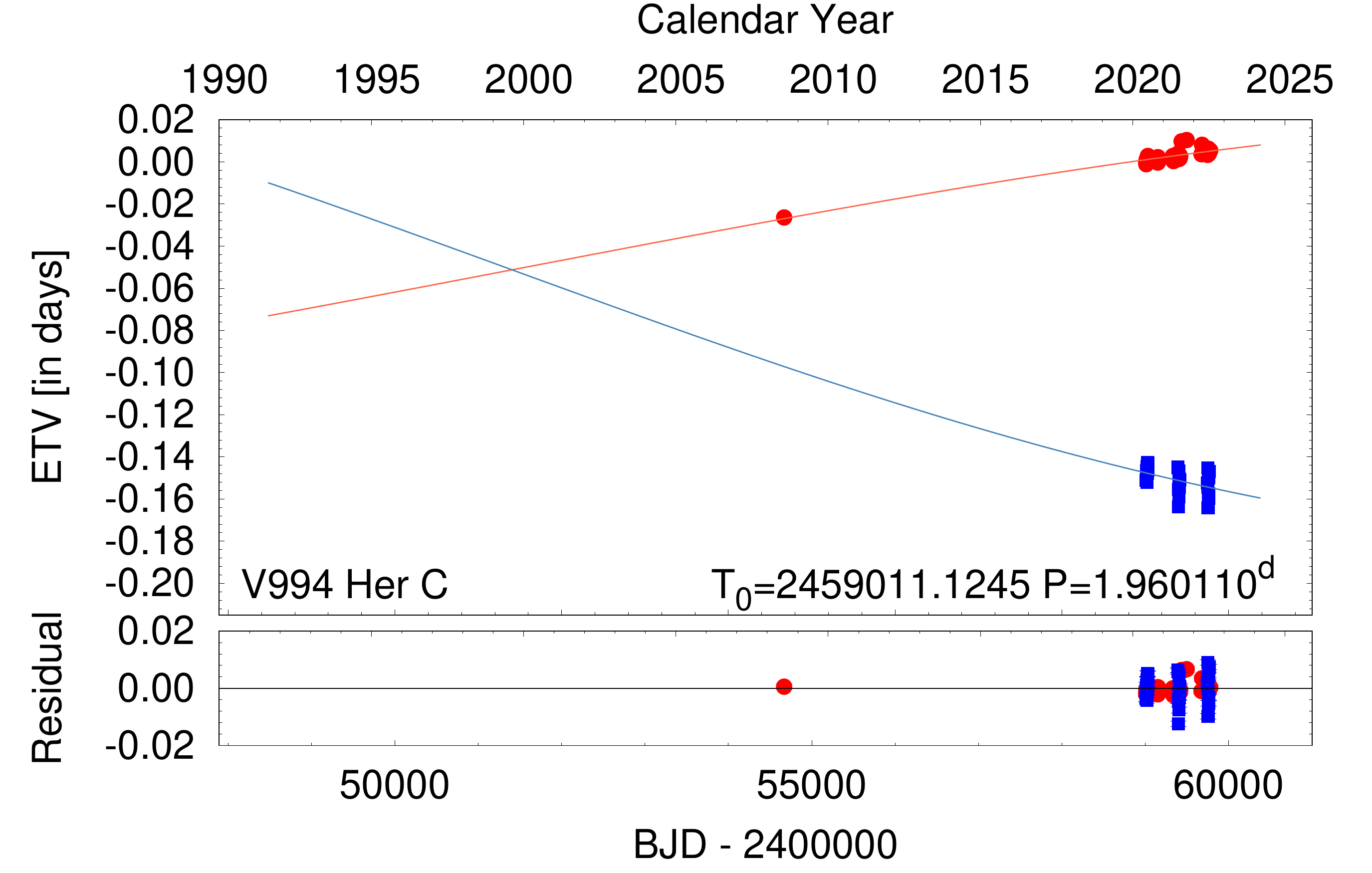} 
\caption{Eclipse timing variations (ETVs) of V994 Her collected over the past three decades, with
the \TESS\ eclipses included. The top, middle and bottom panels show the ETVs for binary A, B and
C, in that order; the red and blue points denote the primary and secondary eclipses, respectively.
The red and blue curves are the photodynamical fitting models. The ``divergence'' of the ETVs for
the primary and secondary eclipses are due to apsidal motion.} \label{fig:ETV}
\end{center}
\end{figure}

All the previously-unpublished eclipse times are given in Table \ref{tbl:ecl-times}. The minima
presented in this work for the first time, as well as the previously-published ones, were used for
a final fit of the data over the whole interval (covering more than 30 years now). This is shown
in Figure \ref{fig:ETV} for all three pairs.

\subsection{Other Catalogs}
We queried the WDS Catalog for archival data on V994 Her and its nearby visual companion
(TIC\,1685970000), with measurements spanning from 1831 to 2015. This data consisted of position
angles and separations for the system between these years. Additionally, we calculated the
position angle and separation for the visual double using data from \Gaia\ DR3 \citep{gaiadr3}.
These data were used in Section \ref{VisualOrbit} to investigate whether or not the visual double
star is gravitationally bound.

\subsection{Speckle observation}
\label{ref:spec_obs} V994 Her was observed on 10 May 2022 using the, `Alopeke speckle
interferometric imager mounted on the Gemini North 8-m telescope \citep{2021FrASS...8..138S}.
`Alopeke provides simultaneous speckle imaging in two bands (562 nm and 832 nm), with output data
products including a reconstructed image and derived parameters for any detected close companion
stars. Three sets of $1000 \times 0.06$~sec exposures were collected; these underwent Fourier
analysis in the standard reduction pipeline \citep{2011AJ....142...19H}.

\begin{figure}
\begin{center}
\includegraphics[width=0.91 \columnwidth]{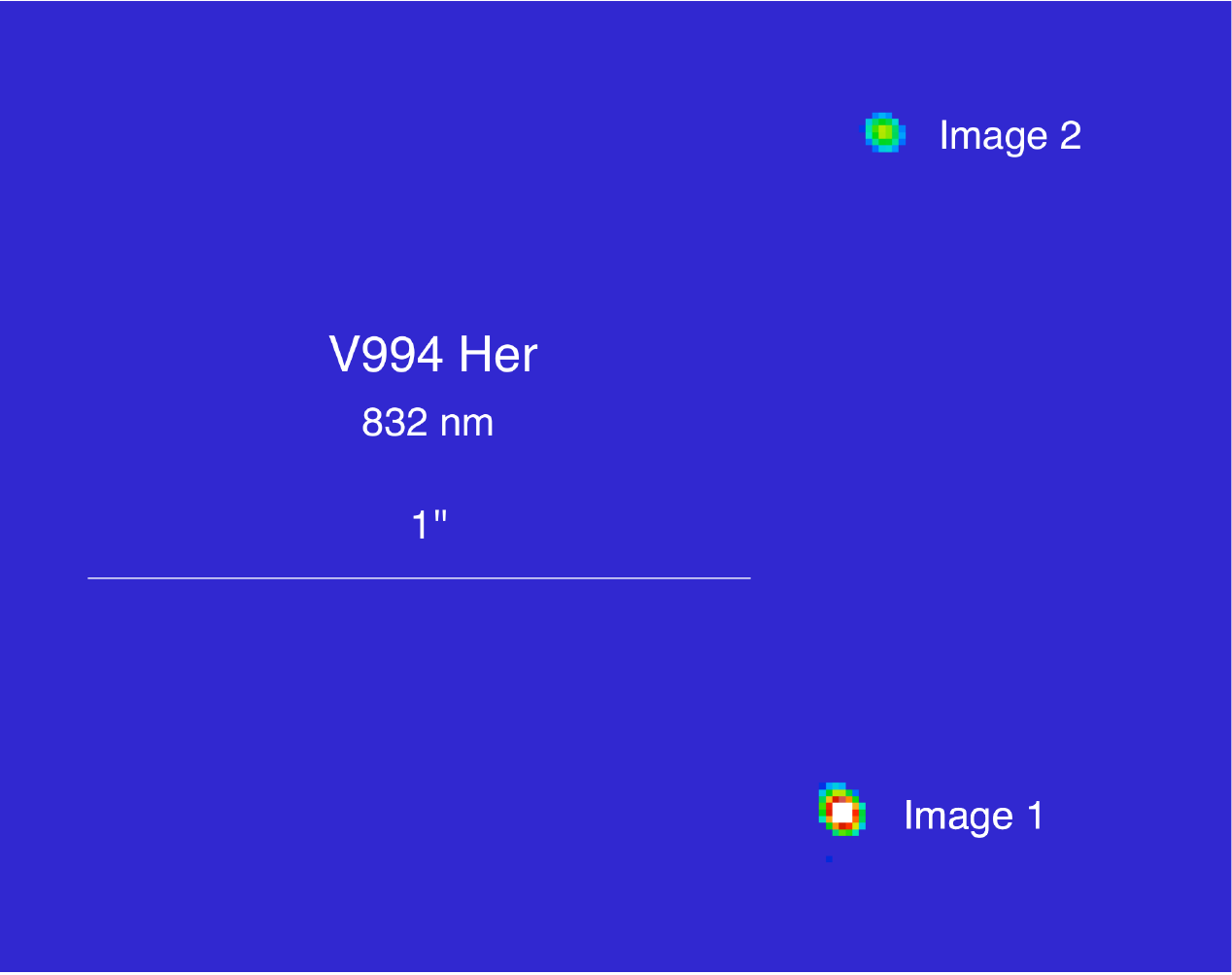} \vglue0.1cm
\includegraphics[width=0.91 \columnwidth]{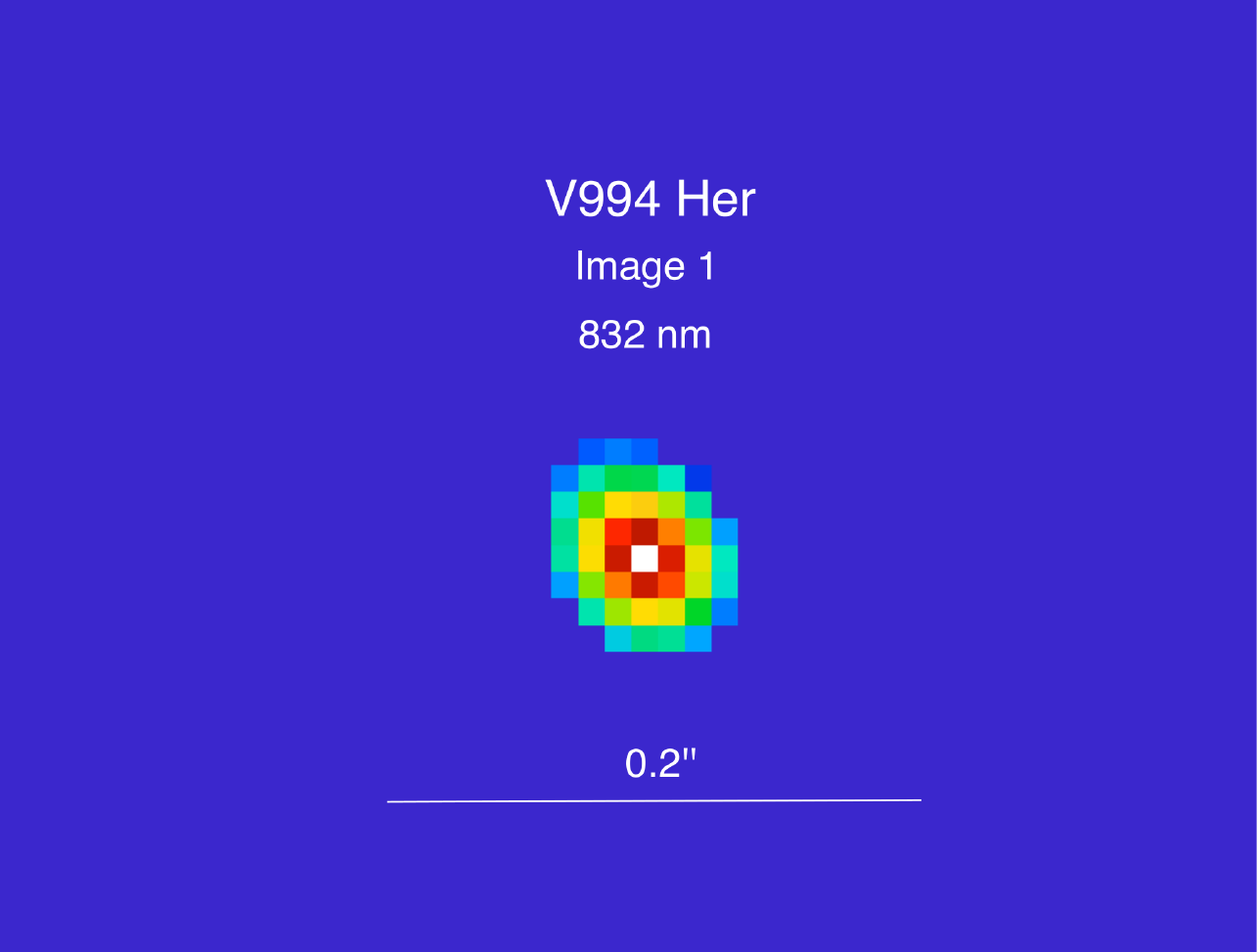} \vglue0.05cm \hglue-0.3cm
\includegraphics[width=0.99 \columnwidth]{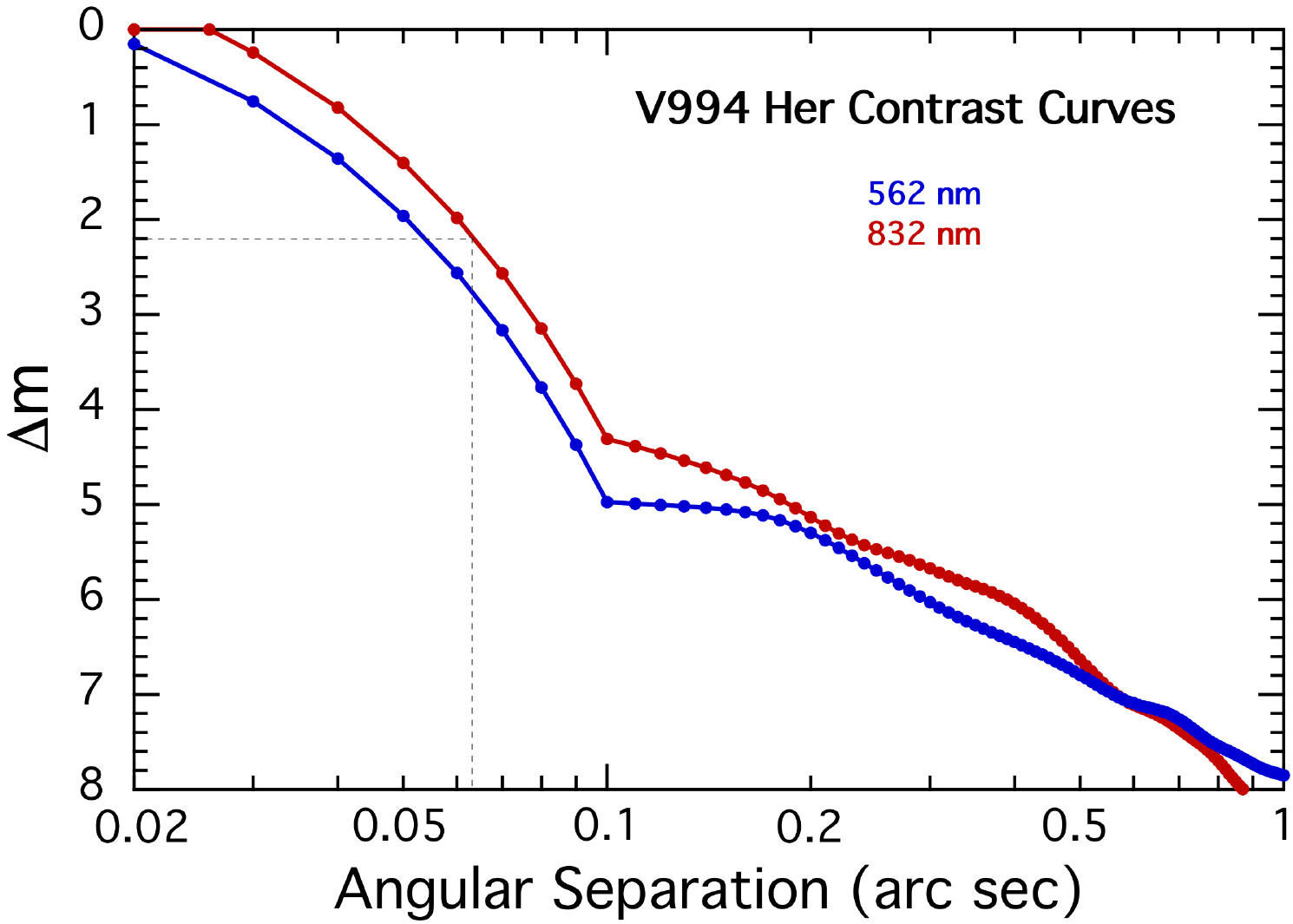}
\caption{Speckle imaging of V994 Her. North is up and East is to the left. {\it Top panel}: 832 nm
image of a $1.5'' \times 1.5''$ region near V994 Her.  We define the brighter feature to the South
as `Image 1,' which contains binaries A and B.  We label the $\sim$1.7 magnitude fainter object to
the North as `Image 2'.  {\it Middle panel}: Same as top panel but zoomed in around Image 1.  Each
pixel is $0.01''$ in size.  {\it Bottom panel}: 5-$\sigma$ confidence level contrast curves
(obtained at 562 nm and 832 nm).  The image spans angular scales from the diffraction limit, near
20 mas, out to $\sim$ $1''$, the approximate end of speckle coherence. The dotted black lines mark
the detectable separation distance of $\sim$ $0.06''$ of a source that is 2.2 magnitudes fainter
than Image 1 itself (i.e., the approximate brightness of binary C).} \label{fig:speckle}
\end{center}
\end{figure}

Figure \ref{fig:speckle} shows the image around V994 Her, with a bright component to the South
(hereafter `Image 1') which hosts binaries A and B.  The fainter image to the North (hereafter
`Image 2') is 1.06$''$ away, and we believe that this image hosts binary C, as discussed in
Section \ref{VisualOrbit}.  At the 290~pc distance to V994~Her (derived from the {\it Gaia} DR3
parallax), this corresponds to a spatial separation of $\sim 307$~au. The middle panel shows a
zoom-in around Image 1, revealing that there are no resolved components within it, down to a
limiting resolution of $\lesssim 0.1''$.  Both the image of the quadruple system to the South
(Image 1), and binary C (likely residing in Image 2) remain unresolved into their component
parts---respectively, either the A and B binaries, and the primary and secondary star in binary
C---as these components are separated on the sky by less than our 20~mas nominal angular
resolution. This value is the Gemini optical diffraction limit when Nyquist sampled with $2 \times
0.01''$ pixels. Our derived 5-$\sigma$ contrast curves for this observation, for both the 562~nm
and 832~nm images, are shown in the bottom panel.  These curves will be further discussed in
Sect.~\ref{VisualOrbit} as we attempt to rule out the possibility that binary C might actually be
located in Image 1.

The system properties gleaned from the speckle observations are summarized in Table
\ref{tbl:speckle}.

\begin{table}
\centering \caption{System parameters derived from speckle imaging$^a$.}
\begin{tabular}{lc}
\hline \hline
 Parameter & Value \\
\hline
obs.~date [JD] & $2459710.018  \pm 0.001$ \\
position angle [deg] & $357.5 \pm 0.5$ \\
separation [arc sec] & $1.06 \pm 0.01$  \\
$\delta V$ [mag]$^b$ &  $1.89 \pm 0.2$ \\
$\delta I$ [mag]$^b$ &  $1.67 \pm 0.2$ \\
\hline
\label{tbl:speckle}  
\end{tabular}

\textit{Notes:}  (a) Observations made at 562 nm and 832 nm with the `Alopeke speckle
interferometric imager mounted on the Gemini North 8-m telescope \citep{2021FrASS...8..138S}. (b)
Difference in magnitude between Image 1 (containing binaries A and B) and Image 2 (likely hosting
binary C).

\end{table}


\section{Photodynamical Modeling} \label{sec:modelling}

We carried out a joint photodynamical modeling in which we combined the three sectors of \TESS\
data alongside the 2007 $V$ band Baja light curves. As part of this, we also modeled the eclipse
timing variation (ETV) curves of all three binaries, the radial velocity (RV) points of binaries A
and B obtained by \citet{2008MNRAS.389.1630L}, and the net stellar spectral energy distributions
(SEDs). To prepare for this analysis, we improved the {\sc Lightcurvefactory} software package to
allow it to handle hierarchical configurations of (2+2)+2 stars in their entirety. Specifically,
the updated code calculates the revolutions of the six bodies on their three inner orbits, the
middle orbit (i.e., of the quadruple), and the outer orbit (i.e., of the sextuple). All five
orbits may be considered either to be purely Keplerian or, for tight systems, {\sc
Lightcurvefactory} is able to take into account the mutual perturbations of the constituent stars
with numerical integration of the orbital motions. Moreover, any combinations of two-body or
multiple-body eclipses are also considered. The updated code does not require the disentangling of
the three eclipsing binary light curves; rather, they can be modeled in their observed, blended
form (e.g., as shown in Figure \ref{fig:rawLC}). Apart from this improvement, the software package
is functionally identical to that described in previous work (see, e.g.,
\citealt{borkovitsetal2019,borkovitsetal2021}).

For the specific case of V994~Her, we find that binaries A and B form a relatively wide 2+2
quadruple system ($P_\mathrm{A-B}/P_\mathrm{B}>P_\mathrm{A-B}/P_\mathrm{A}>500$). As a result, the
gravitational perturbations of the binary components in the 2+2 quadruple are small and can be
described by simple Keplerian orbits. Therefore, we use a simple analytic Keplerian formalism in
order to calculate the stellar positions at any given time, with the slight empirical modification
of considering, for all three binaries, a constant apsidal advance rate (${\dot
\omega}_\mathrm{A,B,C}$) and reference values for the argument of periastron
($\omega_\mathrm{A,B,C}$) at a specific epoch. A physical interpretation of the apsidal motion is
discussed in Section~\ref{APSmotion}.

In our joint photodynamical analysis, we optimized the following parameters using a Markov Chain
Monte Carlo (MCMC) method:

\begin{itemize}
\item[(i)] Orbit-related parameters:
\begin{itemize}
    \item For all four orbits (three eclipsing pairs and the quadruple A-B): The components of the eccentricity vectors at epoch $t_0$: $(e\sin\omega)_\mathrm{A,B,C,A-B}$, $(e\cos\omega)_\mathrm{A,B,C,A-B}$, and the inclinations relative to the plane of the sky: $i_\mathrm{A}$, $i_\mathrm{B}$, $i_\mathrm{C}$, $i_\mathrm{A-B}$.
    \item For the A--B orbit: the period $P_\mathrm{A-B}$ and the periastron passage time $\tau_\mathrm{A-B}$.
    \item For the three eclipsing pairs: the (constant) apsidal advance rates: ${\dot \omega}_\mathrm{A,B,C}$.
\end{itemize}

\item[(ii)] Stellar parameters:
\begin{itemize}
    \item Six mass-related parameters: the masses of the primaries ($m_\mathrm{Aa,Ba,Ca}$), and the mass ratios of the three EBs ($q_\mathrm{A,B,C}$),
    \item The metallicity of the system ([$M/H$]),
    \item The (logarithmic) age of the six coeval stars ($\log\tau$),
    \item The interstellar reddening $E(B-V)$, and
    \item The ``extra light'' contamination ($\ell$) parameters.
\end{itemize}
\end{itemize}

A couple of other parameters were constrained instead of being adjusted or held constant during
our analyses:
\begin{itemize}
\item[(i)] Orbits:
\begin{itemize}
    \item The sidereal orbital periods of the inner binaries ($P_\mathrm{A,B,C}$) and their respective orbital phases (derived using the time of an arbitrary primary eclipse) were constrained internally through the ETV curves.
    \item The systemic radial velocity of the whole sextuple system ($\gamma$) is calculated a posteriori at the end of each trial step by minimizing the value of $\chi^2_\mathrm{RV}$.
\end{itemize}
Note that the (2+2)+2 mode of {\sc Lightcurvefactory} requires the orbital elements of the outermost (AB-C) orbit. In the present case, this orbit is completely unknown. We do know, however, that it must be so wide that we do not expect any observable variations in the positions of the six stars arising from their motion along this orbit. Thus, we chose the elements of this outmost orbit arbitrarily; we use a circular orbit seen face-on with a period of $\sim 30$\,kyr with its parameters kept fixed. \\
\item[(ii)] Stars:
\begin{itemize}
    \item The radii and temperatures of the six stars were calculated with the use of three linear interpolations from the precomputed 3D \texttt{PARSEC} grids (the dimensions were metallicity, logarithmic age, and stellar mass).
    \item The distance of the system (needed for the SED fitting) was calculated a posteriori at the end of each trial step, by minimizing the value of $\chi^2_\mathrm{SED}$. For a detailed explanation of this process, see \citet{borkovitsetal2020}.
\end{itemize}
The atmospheric parameters of the stars were handled in a similar manner as in our previous
photodynamical studies. We utilized a logarithmic limb-darkening law \citep{1970AJ.....75..175K},
for which the passband-dependent linear and non-linear coefficients were interpolated in each
trial step via the tables from the original version of the {\tt Phoebe} software \citep{Phoebe}.
We set the (constant) gravity darkening exponents for five radiative stars to $\beta=1$; for the
coolest, solar-like component Cb, however, we used $\beta=0.32$, which is in line with the
classical model of \citet{lucy67} and is valid for convective stars.
\end{itemize}

Prior to conducting our analysis, we performed some further preparatory steps on the light curves.
First, after disentangling the three eclipsing binaries as in Section~\ref{sec:disentanglement},
we found that the residual \TESS\ light curves contained oscillations with an amplitude of
approximately $2$\% and a characteristic period of $(P_\mathrm{A}-P_\mathrm{B})$. We removed this
oscillation from the light curves before performing the full photodynamical analysis by
subtracting the final residual light curve of the iterative disentanglement process (which
contained this periodic variability) from the original \TESS\ light curves for each sector.
Second, for the sake of equal sampling across sectors, we binned the 2-min sector 26 \TESS\ light
curve to 10-min bins, identical to the cadence time of the sector 40 and 53 FFI light curves. We
also binned the 2007 Baja photometry to 10-min bins. Third, we noticed that the eclipse depths of
all three binaries in sector 26 were deeper by a few percent than the corresponding eclipses in
the sectors 40 and 53 data; the depths in the latter two sectors were similar. As a result, we
assume that in sector 26, the ratio of contaminating light is somewhat lower than in the other two
sectors. Thus, for the sector 26 light curve, we adjusted the amount of contaminating light
independent of sectors 40 and 53. The effect of slightly different flux contamination for the same
star in data from different \TESS\ sectors has been studied previously in the literature (see,
e.g., \citealt{2021ApJS..253...11P}.)

\begin{figure}
    \centering
    \includegraphics[width=1.0\linewidth]{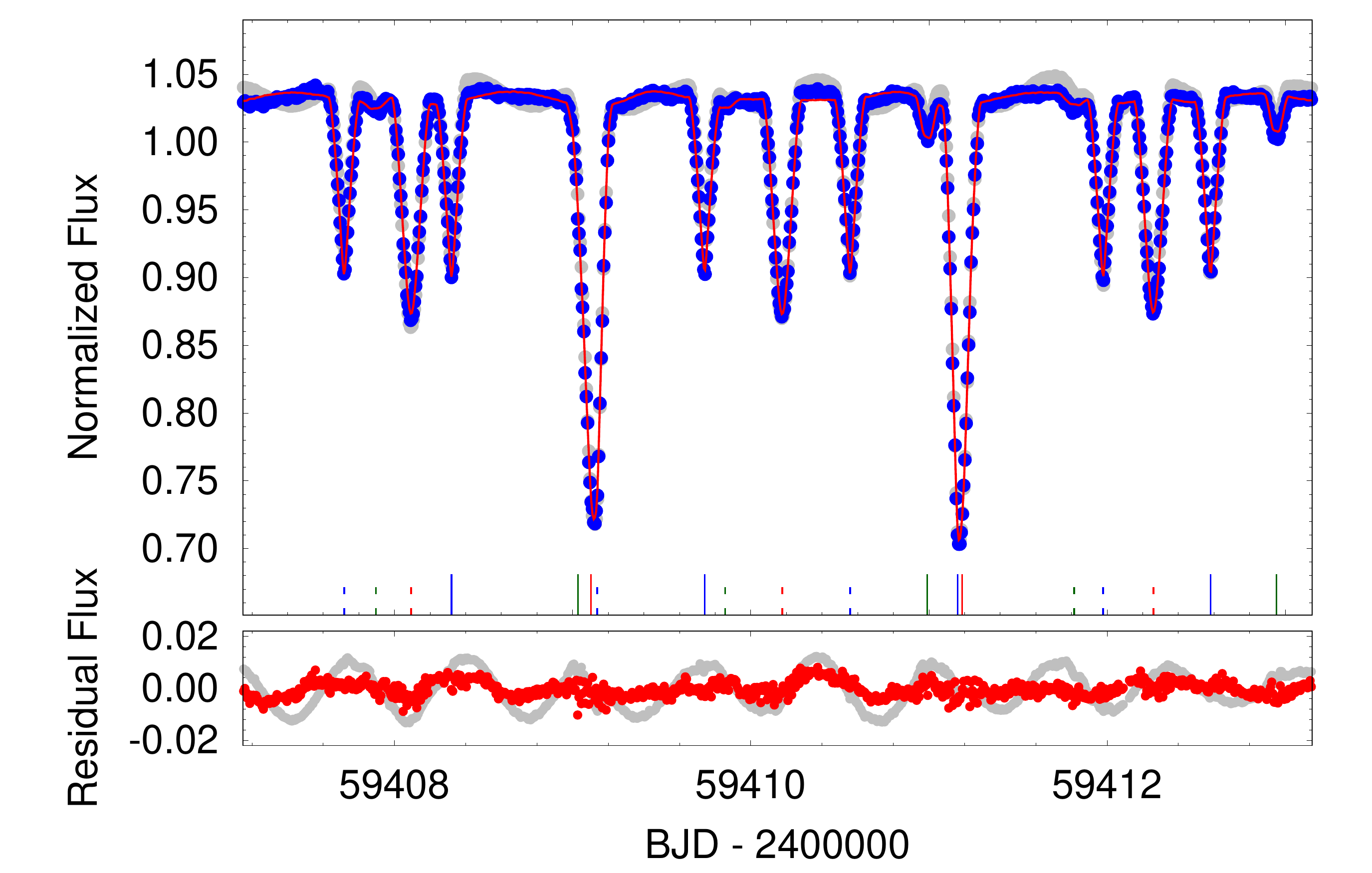}
    \caption{A section of the \TESS\ sector 40 light curve (blue points) after the  removal of the oscillations with period $(P_{\rm A} - P_{\rm B})$, plotted with the photodynamically fitted model light curve (red). We also plot the original light curve, including the oscillations with period $P_A - P_B$, using gray points. The short vertical solid and dashed lines along the x-axis, colored red, blue and green, denote the mid-eclipse times of the primary and secondary eclipses of binaries A, B and C, respectively. The residual curves are shown below the main light curve.}
   \label{fig:TESSlcmodels}
\end{figure}

The median values and their 1-$\sigma$ uncertainties (derived from the MCMC calculation) for the
orbital and physical parameters of the sextuple system, as well as some derived quantities, are
tabulated in Table~\ref{tbl:simlightcurve}. Furthermore, a comparison of the observed and model
light curves are plotted in Figure~\ref{fig:TESSlcmodels}, while a similar comparison for the ETV
curves is shown in Figure~\ref{fig:ETV}. Note that Table~\ref{tbl:simlightcurve} presents the
absolute physical parameters for the C binary and both its components, despite the fact that we do
not have any directly observed radial velocities for this pair. However, we have the RVs of both
the A and B binaries as well as changes of the RVs on their mutual orbit. We can consequently
derive the properties of this binary by using the {\sc Lightcurvefactory} code to combine the
light curve modeling of the C binary and the SED of the overall system.

\begin{table*}
\centering \caption{Median values of the parameters from the joint spectro-photodynamical analysis
of (i) all three EB light curves, (ii) both sets of radial velocities from the SB2 (i.e., the
quadruple consisting of binaries A and B), (iii) all three sets of ETVs, and (iv) joint SED and
PARSEC evolutionary tracks.} \scalebox{0.84}{\begin{tabular}{lccccccc} \hline \hline Parameter &
\multicolumn{2}{c}{Binary A} & \multicolumn{2}{c}{Binary B} & \multicolumn{2}{c}{Binary C} & A--B orbit \\
\hline $P_\mathrm{a}$ [days]                       &
\multicolumn{2}{c}{$2.0832039_{-0.0000039}^{+0.0000042}$ }
& \multicolumn{2}{c}{$1.4200981_{-0.00000040}^{+0.0000033}$} & \multicolumn{2}{c}{$1.9601064_{-0.0000018}^{+0.0000018}$}  & $1062.3_{-2.4}^{+2.8}$ \\
a [$R_\odot$] & \multicolumn{2}{c}{$11.85_{-0.11}^{+0.17}$} & \multicolumn{2}{c}{$8.30_{-0.05}^{+0.07}$} & \multicolumn{2}{c}{$9.45_{-0.23}^{+0.17}$} & $910_{-8}^{+12}$ \\
$i^a$ [deg]  & \multicolumn{2}{c}{$84.66_{-0.33}^{+0.20}$} & \multicolumn{2}{c}{$89.19_{-0.62}^{+0.63}$} & \multicolumn{2}{c}{$80.42_{-0.99}^{+2.39}$} & $83_{-5}^{+4}$ \\
$e$ & \multicolumn{2}{c}{$0.0276_{-0.0010}^{+0.0010}$} & \multicolumn{2}{c}{$0.1186_{-0.0007}^{+0.0007}$} & \multicolumn{2}{c}{$0.1893_{-0.0046}^{+0.0052}$} & $0.687_{-0.037}^{+0.050}$ \\
$\omega$ [deg]  & \multicolumn{2}{c}{$208.4_{-4.3}^{+3.4}$} & \multicolumn{2}{c}{$174.7_{-2.7}^{+2.4}$} & \multicolumn{2}{c}{$314.0_{-2.3}^{+3.4}$} & $59.6_{-3.9}^{+3.6}$ \\
$\dot\omega$ [deg/yr] & \multicolumn{2}{c}{$1.86_{-0.55}^{+0.42}$} & \multicolumn{2}{c}{$3.66_{-0.16}^{+0.17}$} & \multicolumn{2}{c}{$1.68_{-0.24}^{+0.24}$} & $-$ \\
$\tau$ [BJD - 2\,400\,000] & \multicolumn{2}{c}{$59\,010.855_{-0.025}^{+0.020}$} & \multicolumn{2}{c}{$59\,010.393_{-0.011}^{+0.009}$} & \multicolumn{2}{c}{$59\,009.306_{-0.007}^{+0.007}$} & $58\,166.8_{-9.3}^{+13.6}$\\
$t_{\rm prim~eclipse}$ [BJD - 2\,400\,000] & \multicolumn{2}{c}{$59\,011.1821_{-0.0019}^{+0.0014}$} & \multicolumn{2}{c}{$59\,010.7324_{-0.0021}^{+0.0028}$} & \multicolumn{2}{c}{$59\,011.1246_{-0.0004}^{+0.0004}$} & $-$\\
q ($=m_2/m_1$) & \multicolumn{2}{c}{$0.757_{-0.009}^{+0.008}$} & \multicolumn{2}{c}{$1.009_{-0.009}^{+0.009}$} & \multicolumn{2}{c}{$0.583_{-0.074}^{+0.104}$} & $0.738_{-0.013}^{+0.017}$ \\
$K_\mathrm{pri}$ [km\,s$^{-1}$]  & \multicolumn{2}{c}{$124_{-2}^{+2}$} & \multicolumn{2}{c}{$150_{-1}^{+1}$} & \multicolumn{2}{c}{$90_{-8}^{+11}$} & $25_{-1}^{+2}$\\
$K_\mathrm{sec}$ [km\,s$^{-1}$]  & \multicolumn{2}{c}{$163_{-2}^{+2}$} & \multicolumn{2}{c}{$148_{-1}^{+1}$} & \multicolumn{2}{c}{$154_{-7}^{+8}$} & $34_{-2}^{+3}$\\
$\gamma$ [km/s] & \multicolumn{2}{c}{$-$} & \multicolumn{2}{c}{$-$} & \multicolumn{2}{c}{$-$} & $-38.7_{-0.3}^{+0.3}$\\
\hline
individual stars & Aa & Ab & Ba & Bb & Ca & Cb \\
\hline
Relative Quantities: & \\
\hline
fractional radius$^b$ [$R/a$]    & $0.1785_{-0.0026}^{+0.0022}$ & $0.1452_{-0.0015}^{+0.0013}$  & $0.1895_{-0.0019}^{+0.0019}$ & $0.1903_{-0.0017}^{+0.0018}$ & $0.1648_{-0.0039}^{+0.0038}$ & $0.1016_{-0.0106}^{+0.0198}$\\
fractional luminosity in $TESS$-band&  $0.3822_{-0.0143}^{+0.0099}$          & $0.1847_{-0.0072}^{+0.0075}$           &  $0.1227_{-0.0022}^{+0.0025}$          & $0.1258_{-0.0025}^{+0.0029}$ & $0.1086_{-0.0107}^{+0.0199}$ & $0.0145_{-0.0055}^{+0.0113}$ \\
fractional luminosity in $V$-band    & $0.4115_{-0.0175}^{+0.0169}$          & $0.1909_{-0.0115}^{+0.0098}$           &  $0.1184_{-0.0054}^{+0.0041}$          & $0.1220_{-0.0056}^{+0.0047}$ & $0.1021_{-0.0163}^{+0.0227}$ & $0.0085_{-0.0035}^{+0.0074}$ \\
extra light  [$\ell_{S26}$]    &\multicolumn{6}{c}{$0.0576_{-0.0221}^{+0.0155}$} \\
~~~~~~~~~~~~~[$\ell_{S40,53}$] &\multicolumn{6}{c}{$0.1808_{-0.0215}^{+0.0141}$} \\
~~~~~~~~~~~~~[$\ell_V$]        &\multicolumn{6}{c}{$0.0366_{-0.0256}^{+0.0334}$} \\
\hline
Physical Quantities: &  \\
\hline
$T_{\rm eff}^c$ [K]              & $11890_{-264}^{+310}$  & $9915_{-201}^{+256}$  &  $8832_{-156}^{+158}$  &  $8895_{-170}^{+175}$ & $8514_{-310}^{+514}$ & $5893_{-384}^{+451}$ \\
 mass [$M_\odot$]             & $2.929_{-0.093}^{+0.124}$ & $2.216_{-0.070}^{+0.106}$ & $1.913_{-0.040}^{+0.050}$ & $1.889_{-0.049}^{+0.055}$ & $1.810_{-0.072}^{+0.169}$ & $1.077_{-0.109}^{+0.162}$ \\
radius$^c$ [$R_\odot$]           & $2.118_{-0.030}^{+0.026}$ & $1.721_{-0.031}^{+0.038}$ & $1.572_{-0.019}^{+0.028}$ & $1.579_{-0.018}^{+0.027}$ & $1.544_{-0.035}^{+0.071}$ & $0.961_{-0.115}^{+0.199}$ \\
luminosity$^c$  [$L_\odot$]          & $80.2_{-7.7}^{+10.4}$ & $25.7_{-2.6}^{+3.9}$ & $13.5_{-1.0}^{+1.1}$ & $14.1_{-1.2}^{+1.3}$ & $11.2_{-1.8}^{+4.4}$ & $1.00_{-0.41}^{+0.94}$ \\
~~~~~~~~ [$M_\mathrm{bol}$]      & $0.01_{-0.13}^{+0.11}$ & $1.25_{-0.16}^{+0.12}$ & $1.94_{-0.09}^{+0.09}$ & $1.90_{-0.09}^{+0.10}$ & $2.15_{-0.36}^{+0.19}$ & $4.77_{-0.72}^{+0.57}$ \\
$\log \, g^c$  [cgs]               & $4.252_{-0.013}^{+0.017}$ & $4.312_{-0.005}^{+0.005}$ & $4.321_{-0.007}^{+0.007}$ & $4.321_{-0.007}^{+0.007}$ & $4.317_{-0.007}^{+0.009}$ & $4.503_{-0.103}^{+0.065}$ \\
\hline
$\log$(age) [dex]                &\multicolumn{6}{c}{$7.92_{-0.23}^{+0.12}$} \\
$[M/H]$  [dex]                   &\multicolumn{6}{c}{$0.073_{-0.056}^{+0.042}$} \\
$E(B-V)$ [mag]                   &\multicolumn{6}{c}{$0.050_{-0.014}^{+0.011}$} \\
$(M_V)_\mathrm{tot}^c$             &\multicolumn{6}{c}{$-0.30_{-0.07}^{+0.07}$}
\\
distance [pc]                &\multicolumn{6}{c}{$274_{-6}^{+6}$}\\
\hline
\end{tabular}}
\label{tbl:simlightcurve}
 {\em Notes:} (a) Calculated only from the $\sin i$ terms; (b) Polar radii; (c) Interpolated from the PARSEC isochrones
\end{table*}

\section{The final parameters} \label{sec:discussion}

Our thorough modeling of the system also yields the position of each star in the H-R diagram. Due
to the fact that the age of the system was also taken as a free parameter (under the assumption
that all six stars are coeval), we can characterize its evolutionary state. From the calculated
value of the system's logarithmic age (presented in Table \ref{tbl:simlightcurve}), the system is
rather young and therefore located close to the ZAMS. This is in agreement with the fact that all
of the orbits are slightly eccentric, so the circularization process is still ongoing \citep[see,
e.g.,][]{1997A&A...318..187C}.

Only a few sextuple systems have well-constrained parameters, including their masses and orbital
elements; as a result, it is not very easy to compare V994~Her with others. Interestingly, the
recent analysis of the sextuple system TIC\,168789840 \citep{powell_sextuple} revealed a vastly
different configuration, wherein all three binary pairs have similar mass ratios. In V994 Her, our
analysis suggests that all three components have very different mass ratios. Perhaps V994 Her is
more similar to the well-known Castor system, which has a similar architecture, with its component
binaries having very different mass ratios (see, e.g., \citealt{Tokovinin2018}).

The whole system is plausibly close to a co-planar configuration, given the inclination angles in
Table \ref{tbl:simlightcurve}. However, to derive its true orbital architecture we would also need
to calculate the values for the longitude of the ascending node $\Omega$. To do so, one would need
to derive a precise interferometric orbit, which is not available to us currently. Once this
information is obtained, we can speculate whether or not the system can exhibit Kozai-Lidov cycles
(\citealt{1962AJ.....67..591K, 1962P&SS....9..719L}); however, these may be halted anyway by rapid
precession of the pericenters of the component binaries \citep[Table~\ref{tbl:simlightcurve}
and][for an example]{v2016}.

\begin{figure}
\begin{center}
\includegraphics[width=0.48\textwidth]{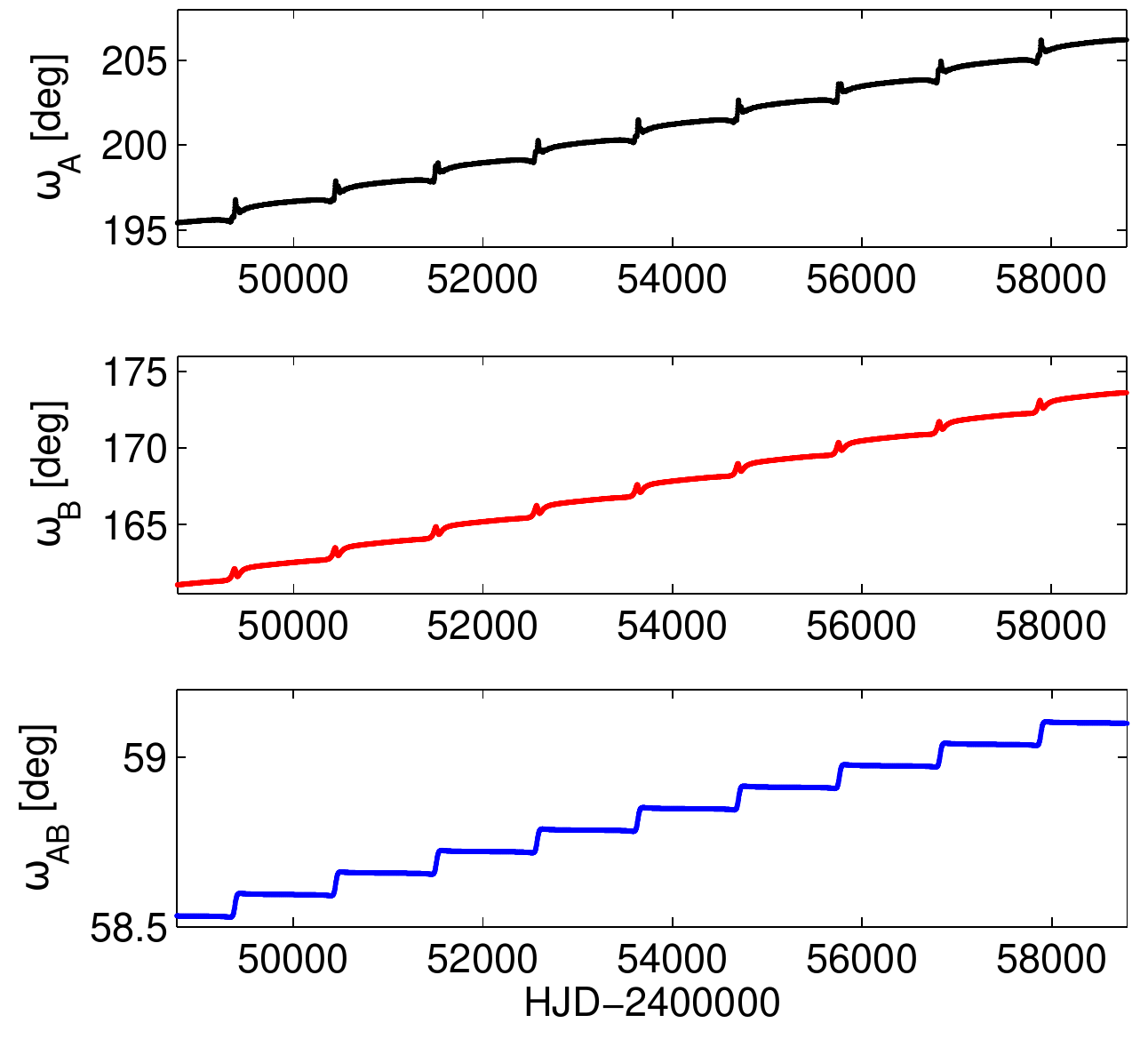}
\caption{Results of the numerical modeling of the orbits of binaries A and B and the quadruple AB.
Here, we show the long-term evolution of the arguments of periastron (without the tidal term).
Pair A is plotted in black; pair B, in red; and their mutual orbit (A-B), in blue. See section
\ref{APSmotion} for details.} \label{fig:Omega}
\end{center}
\end{figure}

\subsection{Apsidal motion} \label{APSmotion}

Given that we have multiple high-precision sets of eclipse times (see Tables
\ref{Tab:V994_Her_A_ToM}, \ref{Tab:V994_Her_B_ToM}, \ref{Tab:V994_Her_C_ToM}, and
\ref{tbl:ecl-times}), and that each binary has an eccentric orbit, we are able to derive apsidal
motion rates for all three pairs in the system. We find that this rate is approximately a few
degrees per year, suggesting that their apsidal advance is not at all negligible. In order to
properly interpret these empirically fitted rates, we first have to subtract any contributions
from the apsidal advance that can be accurately computed.

First, we determine the relativistic contribution to the observed apsidal motion (see, e.g.,
\citealt{2010A&A...519A..57C}). Given the orbital and physical parameters in
Table~\ref{tbl:simlightcurve} we find this effect represents about 10\%, 8\%, and 6\% of the total
for the binaries A, B, and C, respectively. Because these percentages are rather small, we can
consider all three sub-systems of V994~Her as classical apsidal rotators, rather than relativistic
ones.

Because the inner 2+2 component of binaries A and B is not too wide, there also exists a classical
(Newtonian) contribution to their apsidal motion, arising from mutual dynamical perturbations
between A and B. In order to estimate this effect, we ran a simulation using the N-body code
developed by \citet{b2017} and \citet{Broz_2022A&A...666A..24B}. For the sake of definiteness, we
assumed a nearly-coplanar system configuration by imposing identical initial values of the nodal
longitude for both the A and B orbits, with other orbital parameters taken from
Table~\ref{tbl:simlightcurve}. We found that such mutual perturbations in the A and B system
account for another 12\% and 5\% of the total apsidal motion in the respective component (see
Figure \ref{fig:Omega}, which shows these contributions). The binary C is deemed to be distant
enough from the A and B binaries so that we do not provide the classical apsidal contribution in
this case.

With those two effects estimated and subtracted from the total values of the observed apsidal
motions in the A, B and C systems, we can assume that the remainder is attributable to the effect
of the stellar tidal interactions. From these rates of $\dot\omega_\mathrm{tidal}$, one can
usually derive the internal structure constants and compare them with theoretical ones from
stellar evolution models (e.g. \citealt{2004A&A...424..919C}). However, when comparing the results
for pair B (which has the best coverage of its apsidal period, since it has the fastest rate), our
resulting values deviate slightly from the predicted theoretical ones. The tidal contribution to
the apsidal rate was found to be $\dot\omega_\mathrm{tidal,B} = (3.14 \pm 0.20)~\mathrm{ deg/yr}$,
yielding an internal structure constant of $\log k_2= -2.44 \pm 0.05$, while the theoretical
models of \cite{2004A&A...424..919C} suggest that its value should be $-2.36 \pm 0.02$. In order
for the respective error intervals to overlap, one needs to have either larger uncertainties in
the derived parameters, a slightly faster apsidal motion (of about 5\%), subsynchronous rotation
of the component stars (about 20\% slower), or a combination of all three of these effects.
Another way to account for this discrepancy is by using the fact that these stars, found to be
very young, were likely born in a metal-rich environment. Using the stellar evolution models of
\cite{2007A&A...467.1389C} with a higher metallicity (Z=0.04), we find $\log k_2= -2.40 \pm 0.02$,
which is in much better agreement with the observed value of $\log k_2$.

The N-body modelling also allows us to estimate the apsidal advance of the quadruple orbit A-B
(shown in Figure \ref{fig:Omega}). This motion -- accumulating to $\sim 0.6^{\circ}$ over the
interval of available observations -- is orders of magnitude slower when compared to the values
for the A and B systems. However, over the next few decades, when the change will have
cumulatively added up to a few degrees, one can readily detect such movement with newly obtained
data. On the other hand, other effects such as the change in orbital inclination and eclipse depth
would still be negligible on such a timescale.

We also note that the eccentric orbits of the inner binaries are subject to the circularization
effect. From the theory of circularization by \cite{1977A&A....57..383Z} and equations by
\cite{1997A&A...318..187C} the appropriate circularization time scales are of the order of
magnitude longer than the estimated age of the system as resulted from our modelling.



\section{Discussion}
\label{disc}
\subsection{V994 Her and its visual companion} \label{VisualOrbit}

In the prior sections, we  conclusively demonstrated the presence of a third eclipsing binary in
the V994 Her system. Here, we discuss the likelihood that binary C is hosted by Image 2 (fainter
object to the North, as seen in Fig.~\ref{fig:speckle}), as well as the probability that Image 2
is physically bound to Image 1.  If so, this would give the system a (2+2)+2 configuration.

According to the photodynamical fit for the system parameters presented in Table
\ref{tbl:simlightcurve}, binary C has 14\% the light of binaries A+B in the {\it TESS}\ band, and
13\% in $V$ band. That corresponds to magnitude differences of 2.1 and 2.2, respectively, in the
{\it TESS} and V bands.  The bottom panel of Fig.~\ref{fig:speckle} suggests that these contrasts
correspond to being able to resolve two objects within Image 1 (the brighter southerly object)
that are separated by $\gtrsim 0.06''$.  Since both Images 1 and 2 are 290 pc away, the resolvable
physical separation at this magnitude contrast would correspond to 18 au.  The actual semimajor
axis of the binary A and B quadruple, which resides in Image 1, is 4.2 au (see Table
\ref{tbl:simlightcurve}).  It is always possible, of course, that at the time of the speckle
observations, the projected distance between the center of light of binaries A+B and a putative
close orbiting neighbor (i.e., binary C) might inadvertently be very small due to unlucky orbital
phasing.  Let us assume, however, for the sake of argument, that A+B and C are at some nominal
separation on the sky at this particular outer orbital phase when the speckle measurements were
made.  In that case, the outer orbit of C around A+B, within Image 1, would have to have a
semimajor axis of not much more than $\sim 20$~au before it is resolvable.

For a stable triple system (i.e., C stably orbiting A+B) the ratio of semi-major axes must satisfy
\begin{equation}
a_{\rm out} \gtrsim 2.8 \left(\frac{M_{\rm ABC}}{M_{\rm AB}}\right)^{2/5} \frac{(1+e_{\rm
out})^{2/5}}{(1-e_{\rm out})^{6/5}} \,a_{\rm in}\; , \label{eqn:stability}
\end{equation}
where equation (\ref{eqn:stability}) is from \citet{2013ApJ...768...33R}, which in turn is based
on the work of \citet{2001MNRAS.321..398M} and \citet{2008msah.conf...11M}.  If we take as a very
rough estimate that $a_{\rm out} \lesssim 20$~au, and we know that $a_{\rm in} \sim 4.2$~au, then
we find a constraint on $e_{\rm out}$ such that
\begin{equation}
\frac{(1+e_{\rm out})^{2/5}}{(1-e_{\rm out})^{6/5}} \lesssim 1.5\; .
\end{equation}
In turn, this requires that $e_{\rm out} \lesssim 0.25$.  Thus, while this is not an unreasonably
small value for an outer orbital eccentricity, we can see from this exercise, that there is ``not
much room to spare" in trying to fit binary C into an orbit about binaries A+B, all {\it within}
Image 1.  Furthermore, recall that the contrast limits shown in Fig.~\ref{fig:speckle} are
5-$\sigma$ limits.  Thus, we tentatively conclude that binary C, in fact, is hosted by Image 2
(the fainter one to the North).

\begin{figure}
    \centering
    \includegraphics[width=\linewidth]{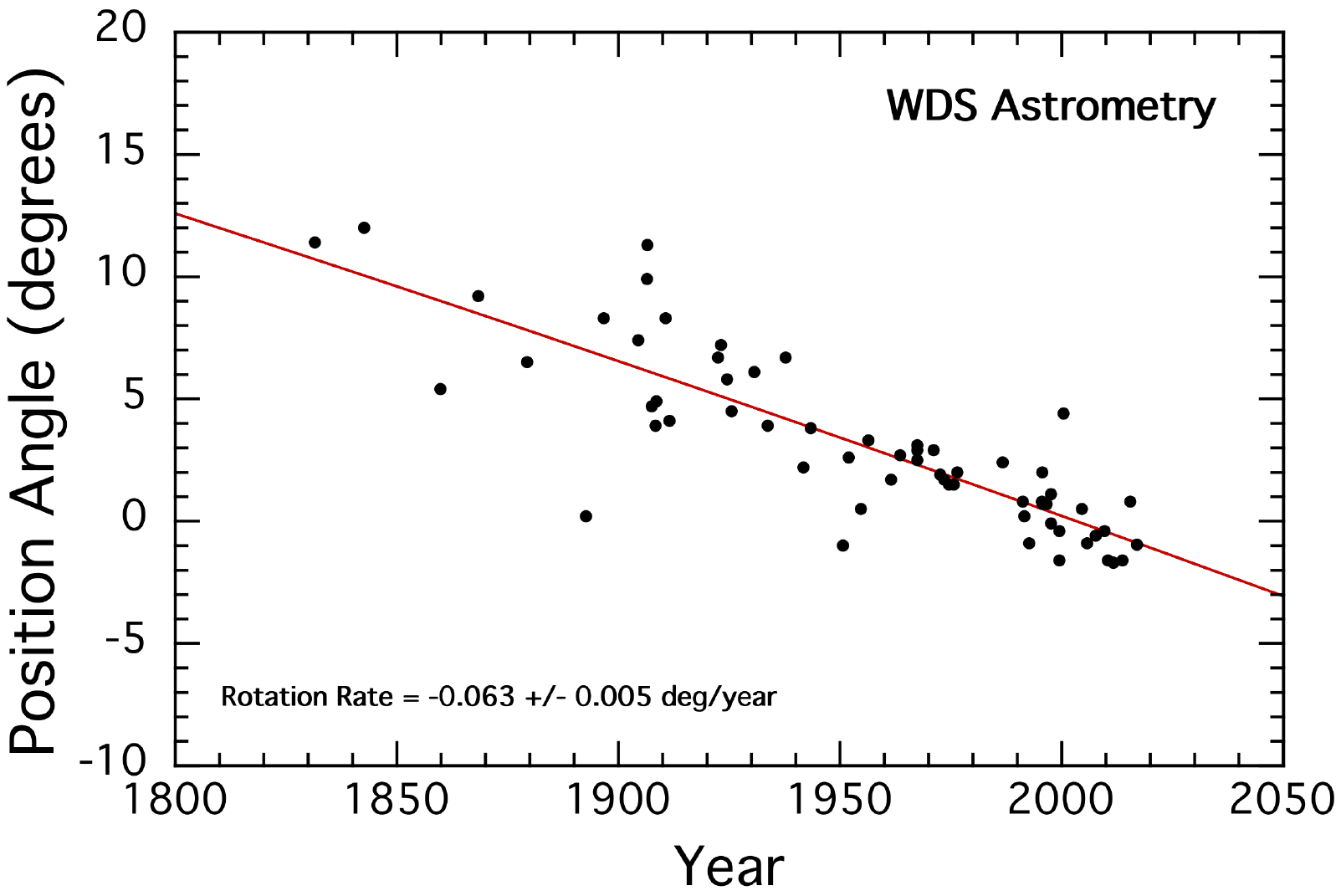}
    \includegraphics[width=\linewidth]{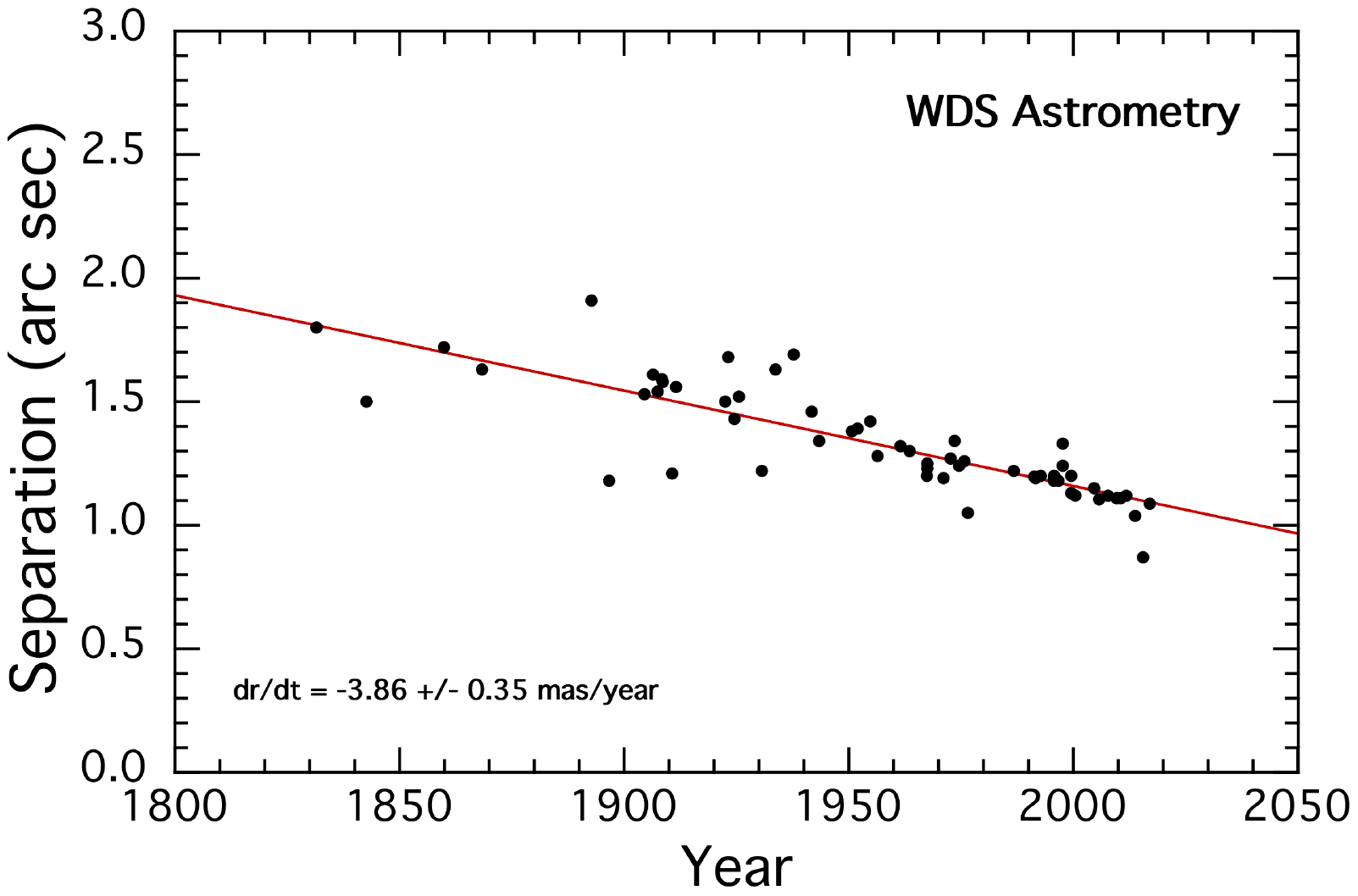}
    \caption{A plot of position angle and separation for the two stars in the visual binary catalogued by the Washington Double Star catalog \citep{2001AJ....122.3466M} at the position of V994 Her.}
    \label{fig:wds-plots}
\end{figure}

We next look at the question of whether Image 2 (likely containing binary C) is physically bound
to Image 1 (hosting binaries A and B).  For this analysis, we have two pieces of kinematic
evidence: (i) the proper motions of Images 1 and 2 from Gaia DR3 \citep{gaiadr3}, and (ii) the
historical astrometric data, spanning 200 years, of the WDS catalog \citep{2001AJ....122.3466M}.
This information is summarized in Table \ref{tbl:kinematics}, and the WDS astrometric data are
plotted in Fig.~\ref{fig:wds-plots}.

\begin{table}
\centering \caption{Observational Kinematics Between Image 1 and Image 2}
\begin{tabular}{lcc}
\hline \hline
 Cartesian motion$^a$ & mas yr$^{-1}$ & km s$^{-1}$ \\
\hline
Gaia  PM RA  & $-0.70 \pm 0.14$ & $-0.96 \pm 0.20$ \\
WDS$^b$ PM RA  & $-1.22 \pm 0.10$  & $-1.68 \pm 0.14$  \\
Gaia  PM Dec & $-4.11 \pm 0.11$ & $-5.66 \pm 0.16$ \\
WDS$^b$ PM Dec & $-3.84 \pm 0.35$  & $-5.29 \pm 0.48$ \\
\hline
angular motion &  --- & --- \\
\hline
WDS$^{b,c}$ $\dot r$  [mas yr$^{-1}$]  & $-3.86 \pm 0.35$ & --- \\
WDS$^{b,c}$ $\dot \theta$ [mrad yr$^{-1}$] & $-1.10 \pm 0.09$ & --- \\
\hline
\label{tbl:kinematics}  
\end{tabular}

\textit{Notes:}  (a) Image 2 value - Image 1. (b) Washington Double Star catalog
\citep{2001AJ....122.3466M}.  (c) These refer to the rate of change in the separation and the
position angle, respectively, and are inferred from the fits shown in Fig.~\ref{fig:wds-plots}.

\end{table}

The two proper motion results (Gaia and WDS) evaluated at the Gaia epoch are in agreement on the
proper motion of the declination (PM Dec) to better than 1 $\sigma$, while the proper motions of
the right ascension (PM RA) differ by 2.9 $\sigma$. We attribute this discrepancy to fitting a
linear function to $\dot \theta$\footnote{$\dot \theta \equiv$ rate of change in the position
angle.} over a 200 year interval. The total relative velocity between Image 1 and Image 2 on the
plane of the sky is in the range 5.55--5.74 km\,s$^{-1}$ depending on whether we choose to use the
WDS or Gaia results, respectively.

To check whether Image 2 is physically bound to Image 1, we take the escape speed to be
\begin{equation}
v_{\rm esc}  \simeq  \sqrt{\frac{2GM_{\rm ABC}}{s}}\; ,
\end{equation}
where $M_{\rm tot}$ is the total mass contained in Image 1 plus Image 2, and $s$ is the
instantaneous (3D) separation of Image 1 and Image 2.  As representative values, we estimate
$M_{\rm ABC} = M_{\rm A}+M_{\rm B}+M_{\rm C} = 11.5$\,M$_{\odot}$ from Table
\ref{tbl:simlightcurve}, and $s \gtrsim 307$~au, where the latter is the physical separation on
the plane of the sky between Image 1 and Image 2.  This leads to an estimate for $v_{\rm esc}
\lesssim  8.2$ km s$^{-1}$.  Since this value is substantially larger than the relative speed of
Image 1 vs.~Image 2 (at least in the plane of the sky), we tentatively take this as strong
evidence that the Image 1 plus Image 2 system is physically bound. However, we remain unsure about
the relative speed and separation in the direction along our line of sight.

Finally, we make another independent argument which also strongly suggests that Image 1 and Image
2 are physically bound. This argument relies on the fact that Image 2 is found so close in the sky
to Image 1, with similar proper motions and distance, and the two are not too dissimilar in
magnitude. We seek to quantify the relative occurrence rate of such a pair of stars. Using Gaia
data, we searched for other stars with similar properties to those of Image 2. In particular, we
looked for stars that have (i) PM RA within an absolute value of 1 mas yr$^{-1}$ of image 1; (ii)
PM Dec with absolute value within 5 mas yr$^{-1}$ of Image 1; (iii) a parallax within absolute
value of 0.2 compared to Image 1; and (iv) having a G magnitude brighter than 9.  When we search
the Gaia database for other stars that satisfy these criteria, we find 13 such stars within
30$^\circ$ of Image 1.  Given that the search area is $\sim$ $10^{10}$ times larger than the area
needed to include Image 2 (at 1$''$ distance from Image 1), we conclude that Image 1, with its
given properties, is not remotely likely to be found there by accident.

The conclusion from the above argument is that either Image 2 is physically bound to Image 1, or
it is comoving with it by virtue of having been born in the same stellar nursery.  The latter
scenario can be ruled out rather readily.  We have seen that the relative speed between Image 1
and Image 2 (on the plane of the sky) is 5.6 km s$^{-1}$.  The age of the system (from Table
\ref{tbl:simlightcurve}) is $\simeq 60$ Myr.  If the two images were unbound and merely
approximately comoving on the sky, then during that time the two images would have drifted apart
in the ensuing 60 Myr by some 300 pc.  This is much, much larger than the current sky separation
of 290 au.  Thus, we conclude that Image 2 is physically bound to Image 1.

Given the above discussion, we believe that the most likely configuration of this sextuple is a
(2+2)+2 system, with the inner quadruple system containing binaries A and B situated in Image 1,
and the third binary (C) in Image 2.  There is also a slim chance, but not yet fully ruled out,
that Image 1 hosts all three binaries, and Image 2 represents a 7th star or yet another binary
(`D').  The most direct ways to prove our most likely scenario (that binary C is in Image 2) is to
(i) check for binary C eclipses in Image 2, and/or (ii) check for RV motions in Image 2 with $P =
1.96$\,d.

\subsection{Outer Orbital Period Distribution}
\label{sect:pout_distr}

Armed with only the relative velocity between Image 1 and Image 2 projected onto the sky, $v_{\rm
sky} \simeq 5.7$ km s$^{-1}$, and the projected separation on the sky, $s_{\rm sky} \simeq
307$~au, we attempt to estimate a probability distribution for the outer orbital period $P_{\rm
out}$ via a Monte Carlo approach. Let $\vec{s}$ and $\vec{v}$ be the full relative position and
velocity vectors between Image 1 and Image 2.  In that case
\begin{eqnarray}
\frac{s_{\rm sky}}{s} & = & \sin \beta_1 \; , \\
\frac{v_{\rm sky}}{v} & = & \sin \beta_2 \; ,
\end{eqnarray}
where $\beta_1$ is the angle between the observer's view direction and $\vec{s}$ and $\beta_2$ is
the angle between the view direction and $\vec{v}$.  If we know nothing about the orientation of
the orbit on the sky, then samples of $\sin \beta$ can be drawn randomly from:
\begin{equation}
\sin \beta = \sqrt{1-\mathcal{R}^2}\; ,
\end{equation}
where $\mathcal{R}$ is a uniformly distributed random number between 0 and 1.  Here, as an
approximation, we treat $\hat{s}$ and $\hat{v}$ as independently and randomly directed with
respect to the observer's view direction.  So, we randomly draw $\beta_1$ and $\beta_2$.

The energy of the outer orbit can now be written as
\begin{equation}
\mathcal{E} = -\frac{GM_1M_2}{s_{\rm sky}}\sin \beta_1+\frac{1}{2} \frac{M_1M_2}{M_{\rm ABC}}
\left(\frac{v_{\rm sky}}{\sin \beta_2}\right)^2 = - \frac{GM_1M_2}{2a_{\rm out}}\; ,
\end{equation}
where $M_1 \equiv M_A+M_B$ and $M_2 \equiv M_C$. This reduces to a simple expression for the
semi-major axis of the outer orbit:
\begin{equation}
 \frac{1}{a_{\rm out}} = \frac{2\sin \beta_1}{s_{\rm sky}}- \left(\frac{v_{\rm sky}}{GM_{\rm ABC} \sin \beta_2}\right)^2 \; .
\end{equation}

\begin{figure}
    \centering
    \includegraphics[width=1.02\columnwidth]{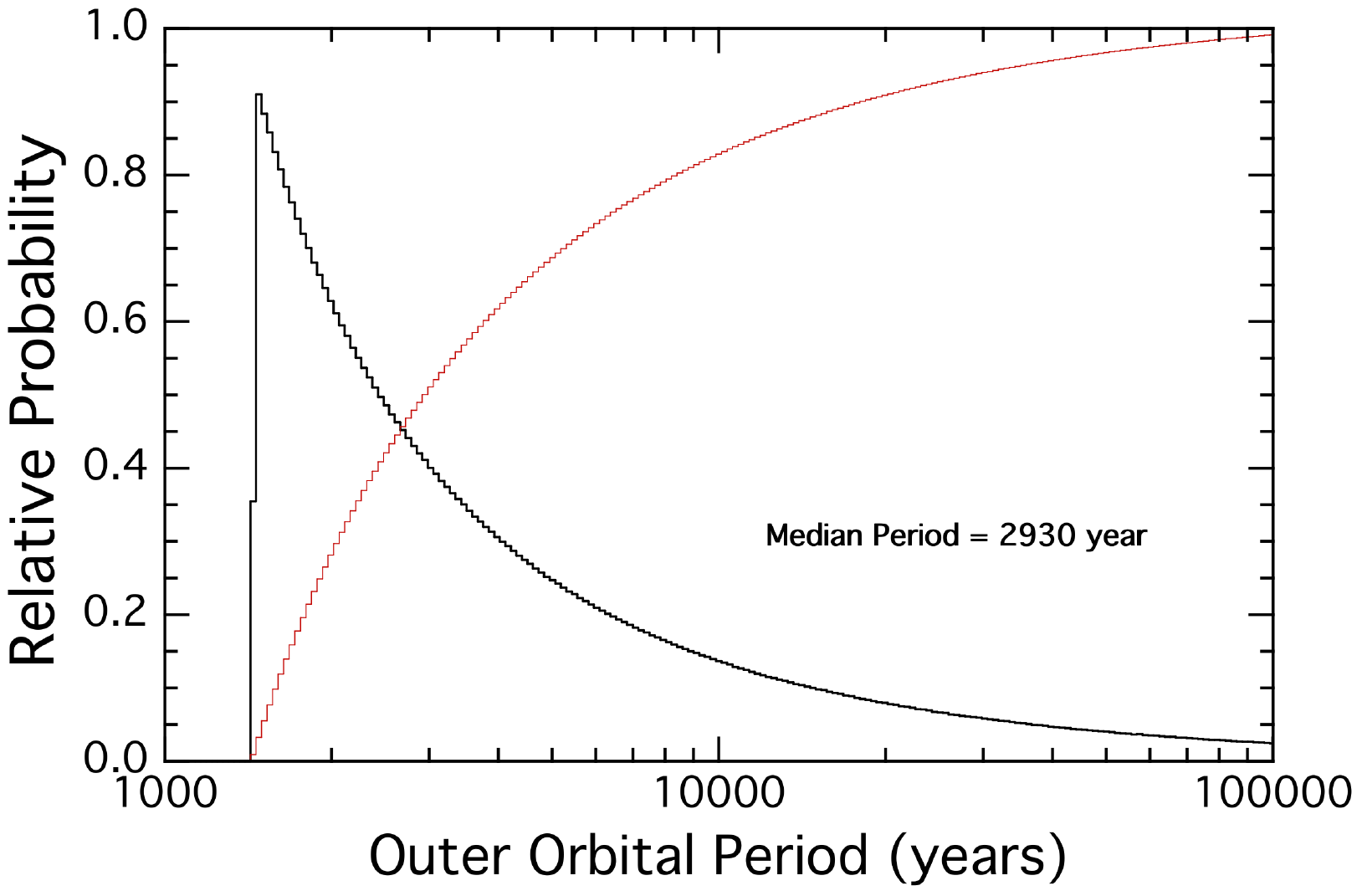}
        \caption{Probability distribution for the outer orbit of the V994 Her system.  Results have been logarithmically binned.  The black histogram is the differential probability distribution, while the red curve is the cumulative distribution.}
    \label{fig:P_out}
\end{figure}

Finally, we make a large number of random draws for $\beta_1$ and $\beta_2$ and, for each
combination, store the realization for the semi-major axis and corresponding $P_{\rm out}$.  The
resultant distribution for $P_{\rm out}$ is shown in Fig.~\ref{fig:P_out}.  The distribution has a
sharp cut at $\simeq 1500$ years, reflecting the fact that the minimum orbital separation is
attained when $\beta_1 = \beta_2 = 90^\circ$, i.e., when the outer orbit is in the plane of the
sky. However, since we do not know the orbital parameters in the line of sight, long orbital
periods are possible, as evidenced by the long tail of the period distribution in
Figure~\ref{fig:P_out}.  The median of the orbital period distribution is close to 3000~years.

We note that our uncertain knowledge of the outer orbit could be greatly improved with a radial
velocity study of Image 2 (which presumably hosts binary C).

\subsection{Possible role of the 3:2 mean motion resonance} \label{res32}

\citet{zetal2019} presented a thorough analysis of stellar quadruple systems with a 2+2
architecture that exhibit eclipses of both components, with binary periods less than 15~days. One
of the interesting population results from this study was the identification of a statistically
significant group of systems having a period ratio close to $3/2$. \citet{zetal2019} speculated
that these systems are either captured in the 3:2 mean motion resonance of the binary periods, or
interacted with this resonance in the recent past and still reside close to it. One of the
consequences for this class of systems would be a possible excitation of the orbital eccentricity
of the binary with the longer period. The V994~Her system was considered in this class by
\citet{zetal2019}. Now, with much more detailed information about V994~Her, we revise its status
with respect to the group of objects that reside or interacted with the 3:2 resonance in the past.

We use an analytical description of the low-order mean motion resonances in 2+2 quadruples by
\citet{t2020} \citep[for completeness, see also][ who discuss the 1:1 mean motion resonant states
in the 2+2 quadruples]{bv2018}. First, it is trivial to check that V994~Her is not currently
located in the resonance since $1-(2P_{\rm A}/3P_{\rm B})\simeq 0.02204$ is too large (it would
need to be three orders of magnitude smaller to be considered to possess this resonance).
\citet{t2020} also discusses sidebands of the pure 3:2 mean motion resonance between $P_{\rm A}$
and $P_{\rm B}$ generated by multiplets of the mean motion frequency $n_{\rm AB}$ of the mutual
orbit. Their importance is typically very small, because the sideband width at frequency
$k\,n_{\rm AB}$ has a multiplicative factor $\propto e^{|k|}_{\rm AB}$ ($k$ is an integer). Here,
$e_{\rm AB}\simeq 0.7$ is a rather large value. However, to account for the three orders of
magnitude in separation between the observed eccentricity and the requirement for resonance, $|k|$
would have to be greater than 30, which is much too large. The system, however, may have crossed
the resonance in the past; this could have contributed to an excitation of the $e_{\rm A}$ value.

In order to place the system into the exact 3:2 mean motion resonance, one would need to (i)
increase $P_{\rm A}$ by $\Delta P_{\rm A}\simeq 0.046943$~day, (ii) decrease $P_{\rm B}$ by
$\Delta P_{\rm B}\simeq 0.031296$~day, or (iii) perform some combination of the two operations.
Additionally, in order to temporarily capture V994~Her in the 3:2 resonance in the past, $P_{\rm
A}$ and $P_{\rm B}$ should have been converging towards each other. In what follows, we shall
discuss an end-member possibility (i) that $P_{\rm B}$ was constant, and $P_{\rm A}$ was evolving
from an initially larger value beyond the resonance condition toward the current value. However,
identical conclusions are obtained for other options, such as keeping $P_{\rm A}$ constant and
$P_{\rm B}$ increasing as in (ii), or their combination.

Using the results from Appendix~C of \citet{t2020}, we note that the putative past capture in the
3:2 resonance puts a severe constraint on the speed by which the period $P_{\rm A}$ decreased. In
particular, denoting the corresponding characteristic timescale $\tau_{\rm A}=P_{\rm A}/{\dot
P}_{\rm A}$ (with ${\dot P}_{\rm A}=-dP_{\rm A}/dt)$, we find that
\begin{equation}
 \tau_{\rm A} \geq K\,(M_{\rm B}/\mu_{\rm B})^{4/3}\,(a_{\rm AB}/a_{\rm B})^{20/3}\,P_{\rm A}
  \; ,
\end{equation}
where $K\simeq 1.74\times 10^{-2}$, $M_{\rm B}$ and $\mu_{\rm B}$ represent the total and reduced
masses of the shorter period binary component, $a_{\rm B}$ and $a_{\rm AB}$ are the semimajor axes
of the B and A-B orbits, and $P_{\rm A}$ is the orbital period of the A binary. Substituting the
values from Table~\ref{tbl:simlightcurve}, we have $\tau_{\rm A}\geq 26$~Gyr. Assuming an
approximately steady decay of the A orbit, we then estimate a minimum time needed to accumulate
the difference $\Delta P_{\rm A}$ between the resonance and the current state
\begin{equation}
 \Delta T_{\rm A} \simeq \frac{\Delta P_{\rm A}}{P_{\rm A}}\,\tau_{\rm A} \geq 580
  \;\;{\rm Myr}  \; .
\end{equation}
This is nearly an order of magnitude longer than the estimated age of the V994~Her system
(Table~\ref{tbl:simlightcurve}). Since other possibilities outlined above lead to the same result,
such as $P_{\rm B}$ drifting toward its current value from an initially smaller value, we may
conclude that the V994~Her system in all likelihood never interacted with the 3:2 mean motion
resonance. Its location near to it might therefore be just coincidental. As a consequence, the
$e_{\rm A}$ value is fully a relic of the initial state, with possible tidal damping. Indeed, the
interaction with the 3:2 resonance would likely not be capable of explaining the significantly
larger $e_{\rm B}$ value. The latter might be excited by interaction with the 2:1 mean motion
resonance between $P_{\rm A}$ and $P_{\rm B}$ values; the location of this resonance, however, is
much too distant from the current system parameters.

\section{Summary and Conclusions}
\label{sec:disc}

In this paper, we have demonstrated that the first-known doubly eclipsing system V994 Herculis is
in fact at least a sextuple system that unambiguously demonstrates three sets of eclipses. Using
\TESS\ and archival data, we have disentangled the light curves of all three binaries in the
system using three different techniques and added new measurements to the O--C diagrams of
binaries A and B. We have also identified the period of the newly-discovered binary C to be 1.9601
days, based on \TESS\ and older ground-based data. Finally, we used archival data from the
Washington Double Star Catalog (spanning over 190 years) alongside parameters from Gaia DR3 in
order to prove that the fainter visual companion on the night sky ($1.1''$ distant) is likely
gravitationally bound to this system and may harbor binary C.

Depending on the nature of the companion star, this could be either a rare (2+2)+2 sextuple star
system---similar to the well-known system Castor, with the same architecture (see, e.g.,
\citealt{2003A&A...402..719S}, and \citealt{Tokovinin2018}). Another possibility is that the
brighter star has six unresolvable stars, and the nearby visual companion is another bound member
of the system, making it even more interesting---a potential septuple (or even
higher-multiplicity) star system. Using additional data, we can more precisely derive the outer
orbit; moreover, updated higher angular-resolution photometry would be able to firmly prove
whether or not the C pair resides in the fainter nearby component. We urge the community to
observe these interesting stars using the tools at their disposal. The high-angular separation
techniques (both in photometry as well as spectroscopy) would be able to prove the true structure
of the system. As the separation of the visual pair on the night sky is slowly decreasing, it may
become increasingly difficult to carry out these observations as time goes on.

\section*{Data availability}

The \TESS\ data underlying this article were accessed using the MAST (Barbara A. Mikulski Archive
for Space Telescopes) Portal
(\url{https://mast.stsci.edu/portal/Mashup/Clients/Mast/Portal.html}). Some of the data were
derived from sources in the public domain; their URLs are provided as footnotes. The derived data
generated in this paper and the code used for the photodynamical analysis will be shared upon
reasonable request to the corresponding author P.Z.

\section*{Acknowledgments}
This paper includes data collected by the \TESS\ mission, specifically as part of GI program
G022062 (PI: A. Pr\v{s}a). Funding for the \TESS\ mission is provided by the NASA Science Mission
directorate. Resources used in this work were provided by the NASA High End Computing (HEC)
Program through the NASA Advanced Supercomputing (NAS) Division at Ames Research Center for the
production of the SPOC data products. Some of the data presented in this paper were obtained from
the Mikulski Archive for Space Telescopes (MAST). STScI is operated by the Association of
Universities for Research in Astronomy, Inc., under NASA contract NAS5-26555. Support for MAST for
non-HST data is provided by the NASA Office of Space Science via grant NNX09AF08G and by other
grants and contracts.

This research has made use of the Washington Double Star Catalog maintained at the U.S. Naval
Observatory, and we thank Rachel Matson for providing archival data on the V994 Her visual double.

This work has used data  from the European Space Agency (ESA)  mission {\it
\Gaia}\footnote{\url{https://www.cosmos.esa.int/gaia}}, processed by  the {\it   Gaia} Data
Processing   and  Analysis   Consortium
(DPAC)\footnote{\url{https://www.cosmos.esa.int/web/gaia/dpac/consortium}}. Funding for the DPAC
is provided  by national institutions, in  particular those participating in the {\it Gaia}
Multilateral Agreement.

Some of the observations in the paper made use of the High-Resolution Imaging instrument `Alopeke,
obtained under Gemini LLP Proposal Number: GN/S-2021A-LP-105. `Alopeke was funded by the NASA
Exoplanet Exploration Program and built at the NASA Ames Research Center by Steve B. Howell, Nic
Scott, Elliott P. Horch, and Emmett Quigley. `Alopeke was mounted on the Gemini North (and/or
South) telescope of the international Gemini Observatory, a program of NSF's NOIR Lab, which is
managed by the Association of Universities for Research in Astronomy (AURA) under a cooperative
agreement with the National Science Foundation, on behalf of the Gemini partnership: the National
Science Foundation (United States), National Research Council (Canada), Agencia Nacional de
Investigaci\'{o}n y Desarrollo (Chile), Ministerio de Ciencia, Tecnolog\'{\i}a e Innovaci\'{o}n
(Argentina), Minist\'{e}rio da Ciencia, Tecnologia, Inova\c{c}\~{o}es e Comunica\c{c}\~{o}es
(Brazil), and Korea Astronomy and Space Science Institute (Republic of Korea).

This work is supported by MEYS (Czech Republic) under the projects MEYS LM2010105, LTT17006 and
EU/MEYS CZ.02.1.01/0.0/0.0/16\_013/0001403 and CZ.02.1.01/0.0/0.0/18\_046/0016007.

M.B. and D.V. were supported by the Czech Science Foundation, grant GA21-11058S.

We  also used the  Simbad  service  operated by  the  Centre des  Donn\'ees Stellaires
(Strasbourg,  France) and the ESO  Science Archive Facility services (data  obtained under request
number 396301).

\bibliography{v994_her}{}
\bibliographystyle{mnras}

\appendix

\section{Table of times of eclipses}

\begin{table*}
\caption{\TESS\ observed times of minima of V994\,Her\,A}
 \label{Tab:V994_Her_A_ToM}
\begin{tabular}{@{}lrllrllrl}
\hline
 Eclipse Time      & Cycle  & std. dev. &  Eclipse Time   & Cycle  & std. dev. &  Eclipse Time  & Cycle  & std. dev. \\
 BJD-2400000 & no.    &   \multicolumn{1}{c}{$(d)$} & BJD-2400000 & no. &   \multicolumn{1}{c}{$(d)$} & BJD-2400000 & no. &   \multicolumn{1}{c}{$(d)$} \\
\hline
59011.208388 &    0.0 & 0.000068 & 59393.510297 &  183.5 & 0.000244 & 59744.520841 &  352.0 & 0.000102 \\
59012.282173 &    0.5 & 0.000038 & 59394.519437 &  184.0 & 0.000103 & 59745.592606 &  352.5 & 0.000108 \\
59013.291644 &    1.0 & 0.000027 & 59395.593429 &  184.5 & 0.000134 & 59746.604119 &  353.0 & 0.000102 \\
59014.365223 &    1.5 & 0.000039 & 59396.602611 &  185.0 & 0.000128 & 59747.675710 &  353.5 & 0.000085 \\
59015.375104 &    2.0 & 0.000028 & 59397.676809 &  185.5 & 0.000113 & 59748.686773 &  354.0 & 0.000100 \\
59016.448767 &    2.5 & 0.000038 & 59398.686003 &  186.0 & 0.000110 & 59749.759375 &  354.5 & 0.000097 \\
59017.458488 &    3.0 & 0.000028 & 59399.760013 &  186.5 & 0.000116 & 59750.770102 &  355.0 & 0.000093 \\
59018.532198 &    3.5 & 0.000044 & 59400.769258 &  187.0 & 0.000105 & 59751.842493 &  355.5 & 0.000101 \\
59019.541632 &    4.0 & 0.000028 & 59401.843411 &  187.5 & 0.000115 & 59752.853041 &  356.0 & 0.000097 \\
59020.615100 &    4.5 & 0.000036 & 59402.852513 &  188.0 & 0.000098 & 59753.926086 &  356.5 & 0.000109 \\
59021.624954 &    5.0 & 0.000030 & 59403.926596 &  188.5 & 0.000173 & 59754.937116 &  357.0 & 0.000153 \\
59023.708034 &    6.0 & 0.000040 & 59407.018862 &  190.0 & 0.000150 & 59757.020019 &  358.0 & 0.000086 \\
59025.791322 &    7.0 & 0.000031 & 59408.093094 &  190.5 & 0.000142 & 59759.103225 &  359.0 & 0.000084 \\
59027.874755 &    8.0 & 0.000029 & 59410.176195 &  191.5 & 0.000125 & 59760.175520 &  359.5 & 0.000103 \\
59028.948286 &    8.5 & 0.000040 & 59411.185709 &  192.0 & 0.000125 & 59761.186560 &  360.0 & 0.000087 \\
59029.957752 &    9.0 & 0.000028 & 59412.259529 &  192.5 & 0.000132 & 59762.259204 &  360.5 & 0.000102 \\
59031.031620 &    9.5 & 0.000041 & 59413.268755 &  193.0 & 0.000107 & 59763.269832 &  361.0 & 0.000082 \\
59032.041094 &   10.0 & 0.000030 & 59414.342947 &  193.5 & 0.000129 & 59764.342122 &  361.5 & 0.000094 \\
59033.114938 &   10.5 & 0.000039 & 59415.352144 &  194.0 & 0.000113 & 59765.353549 &  362.0 & 0.000092 \\
59034.124636 &   11.0 & 0.000032 & 59416.426295 &  194.5 & 0.000120 & 59766.426020 &  362.5 & 0.000132 \\
59391.426417 &  182.5 & 0.004700 & 59417.435270 &  195.0 & 0.000112 & 59767.436725 &  363.0 & 0.000098 \\
59392.436550 &  183.0 & 0.000110 & 59418.509214 &  195.5 & 0.000165 & 59768.508944 &  363.5 & 0.000190 \\
\hline
\end{tabular}
\end{table*}

\begin{table*}
\caption{\TESS\ observed times of minima of V994\,Her\,B}
 \label{Tab:V994_Her_B_ToM}
\begin{tabular}{@{}lrllrllrl}
\hline
 Eclipse Time      & Cycle  & std. dev. &  Eclipse Time   & Cycle  & std. dev. &  Eclipse Time  & Cycle  & std. dev. \\
 BJD-2400000 & no.    &   \multicolumn{1}{c}{$(d)$} & BJD-2400000 & no. &   \multicolumn{1}{c}{$(d)$} & BJD-2400000 & no. &   \multicolumn{1}{c}{$(d)$} \\
\hline
59010.698099 &    0.0 & 0.000046 & 59392.097175 &  268.5 & 0.000104 & 59418.259993 &  287.0 & 0.000104 \\
59011.515257 &    0.5 & 0.000044 & 59392.700110 &  269.0 & 0.000111 & 59744.251213 &  516.5 & 0.000111 \\
59012.118117 &    1.0 & 0.000041 & 59393.517972 &  269.5 & 0.000156 & 59744.855372 &  517.0 & 0.000121 \\
59012.934955 &    1.5 & 0.000038 & 59394.119951 &  270.0 & 0.000103 & 59745.671247 &  517.5 & 0.000131 \\
59013.537998 &    2.0 & 0.000042 & 59394.936970 &  270.5 & 0.000111 & 59746.274246 &  518.0 & 0.000112 \\
59014.355223 &    2.5 & 0.000040 & 59395.539982 &  271.0 & 0.000149 & 59747.092093 &  518.5 & 0.000150 \\
59014.957744 &    3.0 & 0.000046 & 59396.357565 &  271.5 & 0.000083 & 59747.695762 &  519.0 & 0.000105 \\
59015.774811 &    3.5 & 0.000043 & 59396.960021 &  272.0 & 0.000105 & 59748.511739 &  519.5 & 0.000095 \\
59016.377966 &    4.0 & 0.000042 & 59397.776849 &  272.5 & 0.000149 & 59749.114405 &  520.0 & 0.000130 \\
59017.194853 &    4.5 & 0.000039 & 59398.379989 &  273.0 & 0.000117 & 59749.931728 &  520.5 & 0.000125 \\
59017.798052 &    5.0 & 0.000041 & 59399.197749 &  273.5 & 0.000093 & 59750.535692 &  521.0 & 0.000125 \\
59018.614518 &    5.5 & 0.000035 & 59399.800066 &  274.0 & 0.000158 & 59751.351793 &  521.5 & 0.000074 \\
59019.218026 &    6.0 & 0.000041 & 59400.617105 &  274.5 & 0.000128 & 59751.954686 &  522.0 & 0.000141 \\
59020.035181 &    6.5 & 0.000040 & 59401.219664 &  275.0 & 0.000102 & 59752.771888 &  522.5 & 0.000125 \\
59020.638250 &    7.0 & 0.000041 & 59402.037976 &  275.5 & 0.000125 & 59753.376286 &  523.0 & 0.000101 \\
59021.455270 &    7.5 & 0.000042 & 59402.640244 &  276.0 & 0.000103 & 59754.192150 &  523.5 & 0.000128 \\
59022.058251 &    8.0 & 0.000047 & 59403.457233 &  276.5 & 0.000109 & 59754.794952 &  524.0 & 0.000153 \\
59023.478273 &    9.0 & 0.000047 & 59404.059974 &  277.0 & 0.000109 & 59755.612006 &  524.5 & 0.000110 \\
59024.295584 &    9.5 & 0.000043 & 59405.480590 &  278.0 & 0.000133 & 59757.032126 &  525.5 & 0.000142 \\
59024.898039 &   10.0 & 0.000047 & 59406.297671 &  278.5 & 0.000102 & 59757.634921 &  526.0 & 0.000116 \\
59025.715847 &   10.5 & 0.000044 & 59406.900386 &  279.0 & 0.000163 & 59758.452032 &  526.5 & 0.000112 \\
59026.318513 &   11.0 & 0.000048 & 59407.718358 &  279.5 & 0.000116 & 59759.055882 &  527.0 & 0.000183 \\
59027.136039 &   11.5 & 0.000051 & 59408.320651 &  280.0 & 0.000110 & 59759.872291 &  527.5 & 0.000123 \\
59027.738752 &   12.0 & 0.000044 & 59409.137946 &  280.5 & 0.000172 & 59760.475192 &  528.0 & 0.000116 \\
59028.555943 &   12.5 & 0.000043 & 59409.740282 &  281.0 & 0.000106 & 59761.291883 &  528.5 & 0.000129 \\
59029.158935 &   13.0 & 0.000039 & 59410.558443 &  281.5 & 0.000107 & 59761.895816 &  529.0 & 0.000116 \\
59029.975931 &   13.5 & 0.000043 & 59411.160938 &  282.0 & 0.000165 & 59762.712201 &  529.5 & 0.000123 \\
59030.578838 &   14.0 & 0.000042 & 59411.977957 &  282.5 & 0.000112 & 59763.315197 &  530.0 & 0.000126 \\
59031.395642 &   14.5 & 0.000044 & 59412.580171 &  283.0 & 0.000096 & 59764.132083 &  530.5 & 0.000106 \\
59031.998690 &   15.0 & 0.000045 & 59413.397720 &  283.5 & 0.000171 & 59764.735789 &  531.0 & 0.000140 \\
59032.815480 &   15.5 & 0.000042 & 59414.000752 &  284.0 & 0.000117 & 59765.552266 &  531.5 & 0.000117 \\
59033.418734 &   16.0 & 0.000045 & 59414.817875 &  284.5 & 0.000108 & 59766.155117 &  532.0 & 0.000126 \\
59034.235331 &   16.5 & 0.000040 & 59415.420146 &  285.0 & 0.000182 & 59766.972072 &  532.5 & 0.000101 \\
59034.838698 &   17.0 & 0.000044 & 59416.238006 &  285.5 & 0.000118 & 59767.575534 &  533.0 & 0.000120 \\
59390.677560 &  267.5 & 0.000177 & 59416.840945 &  286.0 & 0.000127 & 59768.392581 &  533.5 & 0.000089 \\
59391.279999 &  268.0 & 0.000108 & 59417.658130 &  286.5 & 0.000136 &&& \\
\hline
\end{tabular}
\end{table*}

\begin{table*}
\caption{\TESS\ observed times of minima of V994\,Her\,C}
 \label{Tab:V994_Her_C_ToM}
\begin{tabular}{@{}lrllrllrl}
\hline
 Eclipse Time      & Cycle  & std. dev. &  Eclipse Time   & Cycle  & std. dev. &  Eclipse Time  & Cycle  & std. dev. \\
 BJD-2400000 & no.    &   \multicolumn{1}{c}{$(d)$} & BJD-2400000 & no. &   \multicolumn{1}{c}{$(d)$} & BJD-2400000 & no. &   \multicolumn{1}{c}{$(d)$} \\
\hline
59011.124177 &    0.0 & 0.000157 & 59392.220634 &  194.5 & 0.002012 & 59418.829817 &  208.0 & 0.000596 \\
59011.953279 &    0.5 & 0.000895 & 59393.348483 &  195.0 & 0.000439 & 59744.209501 &  374.0 & 0.001378 \\
59013.083475 &    1.0 & 0.000176 & 59394.181279 &  195.5 & 0.001782 & 59746.169255 &  375.0 & 0.000467 \\
59013.916603 &    1.5 & 0.000925 & 59395.308080 &  196.0 & 0.000453 & 59746.993487 &  375.5 & 0.002726 \\
59015.045185 &    2.0 & 0.000170 & 59396.130606 &  196.5 & 0.002677 & 59748.129163 &  376.0 & 0.000505 \\
59015.878338 &    2.5 & 0.000773 & 59397.270147 &  197.0 & 0.000509 & 59748.960592 &  376.5 & 0.002910 \\
59017.005797 &    3.0 & 0.000173 & 59398.082565 &  197.5 & 0.007348 & 59750.091678 &  377.0 & 0.000459 \\
59017.835692 &    3.5 & 0.000770 & 59399.229029 &  198.0 & 0.000449 & 59750.911553 &  377.5 & 0.016446 \\
59018.965311 &    4.0 & 0.000174 & 59401.187681 &  199.0 & 0.000375 & 59752.050606 &  378.0 & 0.000432 \\
59019.796083 &    4.5 & 0.001014 & 59402.019831 &  199.5 & 0.007591 & 59752.873428 &  378.5 & 0.057992 \\
59020.925121 &    5.0 & 0.000169 & 59403.148862 &  200.0 & 0.000397 & 59754.009831 &  379.0 & 0.000405 \\
59021.753129 &    5.5 & 0.001266 & 59403.972393 &  200.5 & 0.003418 & 59754.822117 &  379.5 & 0.003540 \\
59023.720603 &    6.5 & 0.001604 & 59405.927779 &  201.5 & 0.004661 & 59755.972515 &  380.0 & 0.000694 \\
59024.845884 &    7.0 & 0.000172 & 59407.069016 &  202.0 & 0.000706 & 59757.931960 &  381.0 & 0.000433 \\
59025.682738 &    7.5 & 0.001401 & 59407.894095 &  202.5 & 0.057096 & 59758.756173 &  381.5 & 0.005398 \\
59026.806343 &    8.0 & 0.000171 & 59409.028342 &  203.0 & 0.000656 & 59759.890913 &  382.0 & 0.000483 \\
59027.640483 &    8.5 & 0.000943 & 59409.856410 &  203.5 & 0.002463 & 59760.706877 &  382.5 & 0.010731 \\
59028.767693 &    9.0 & 0.000186 & 59410.989522 &  204.0 & 0.000463 & 59761.850436 &  383.0 & 0.000442 \\
59029.597642 &    9.5 & 0.001409 & 59411.812507 &  204.5 & 0.003274 & 59762.669061 &  383.5 & 0.003683 \\
59030.727133 &   10.0 & 0.000175 & 59412.949612 &  205.0 & 0.000503 & 59763.811861 &  384.0 & 0.000536 \\
59031.559275 &   10.5 & 0.000946 & 59414.908974 &  206.0 & 0.000444 & 59764.636994 &  384.5 & 0.001506 \\
59032.688509 &   11.0 & 0.000180 & 59415.758940 &  206.5 & 0.002578 & 59765.771756 &  385.0 & 0.000524 \\
59033.522849 &   11.5 & 0.001363 & 59416.870114 &  207.0 & 0.000460 & 59767.732579 &  386.0 & 0.000409 \\
59034.648313 &   12.0 & 0.000173 & 59417.696368 &  207.5 & 0.001980 & 59768.560018 &  386.5 & 0.007387 \\
59391.388607 &  194.0 & 0.000580 &&&&& \\
\hline
\end{tabular}
\end{table*}

\begin{table*}
\centering \caption{New, unpublished eclipse times of V994 Her for binaries A, B, and C.}
\begin{tabular}{cccccccccc}
\hline \hline
 Eclipse Time & std. dev.  & Pair  & Type  & Reference/& Eclipse Time & std. dev. & Pair  & Type  & Reference/      \\
 BJD-2400000  & \multicolumn{1}{c}{$(d)$} &[A/B/C]& [P/S] & Observer  & BJD-2400000  & \multicolumn{1}{c}{$(d)$} &[A/B/C]& [P/S] & Observer   \\
\hline
  57843.57013 &  0.00045  &   A  &   S   &   R.U.   &  59802.47556 & 0.00045  &   B  &   S  &   R.U.         \\
  57855.59639 &  0.00073  &   B  &   S   &   R.U.   &  59804.49687 & 0.00048  &   B  &   P  &   R.U.         \\
  57902.45775 &  0.00069  &   B  &   S   &   R.U.   &  59815.34773 & 0.00192  &   A  &   P  &   R.U.         \\
  57916.48823 &  0.00185  &   A  &   S   &   R.U.   &  59816.42734 & 0.00148  &   A  &   S  &   R.U.         \\
  57917.49528 &  0.00029  &   A  &   P   &   R.U.   &  59817.43375 & 0.00030  &   A  &   P  &   R.U.         \\
  57940.40624 &  0.00014  &   A  &   P   &   R.U.   &  59817.43433 & 0.00036  &   A  &   P  &   FRAM         \\
  57946.48532 &  0.00097  &   B  &   S   &   R.U.   &  59818.50532 & 0.00052  &   A  &   S  &   FRAM         \\
  57968.38272 &  0.00099  &   B  &   P   &   R.U.   &  59121.62037 & 0.00054  &   A  &   P  &   G.P.         \\
  58232.53401 &  0.00037  &   B  &   P   &   R.U.   &  52509.78153 & 0.00211  &   B  &   P  &   S.D. \& U.M. \\
  58257.48416 &  0.00075  &   B  &   S   &   R.U.   &  52510.52766 & 0.00651  &   B  &   S  &   S.D. \& U.M. \\
  58290.38397 &  0.00021  &   A  &   P   &   R.U.   &  52702.91006 & 0.00275  &   B  &   P  &   S.D. \& U.M. \\
  58343.53778 &  0.00079  &   A  &   S   &   R.U.   &  52703.66135 & 0.00066  &   B  &   S  &   S.D. \& U.M. \\
  58387.30392 &  0.00032  &   B  &   P   &   R.U.   &  52810.86153 & 0.00098  &   B  &   P  &   S.D. \& U.M. \\
  58565.62065 &  0.00055  &   B  &   S   &   R.U.   &  52811.60611 & 0.00190  &   B  &   S  &   S.D. \& U.M. \\
  58570.62894 &  0.00082  &   A  &   S   &   R.U.   &  53096.26742 & 0.00274  &   B  &   P  &   S.D. \& U.M. \\
  58571.63882 &  0.00069  &   A  &   P   &   R.U.   &  53097.02123 & 0.00193  &   B  &   S  &   S.D. \& U.M. \\
  58593.55169 &  0.00068  &   A  &   S   &   R.U.   &  53171.53199 & 0.00181  &   B  &   P  &   S.D. \& U.M. \\
  58614.50216 &  0.00135  &   B  &   P   &   R.U.   &  53172.28521 & 0.00092  &   B  &   S  &   S.D. \& U.M. \\
  58667.47186 &  0.00029  &   A  &   P   &   R.U.   &  52692.65792 & 0.00048  &   A  &   P  &   S.D. \& U.M. \\
  58689.37993 &  0.00062  &   A  &   S   &   R.U.   &  52715.57549 & 0.00129  &   A  &   P  &   S.D. \& U.M. \\
  58957.55352 &  0.00149  &   B  &   S   &   R.U.   &  53139.55537 & 0.00089  &   A  &   S  &   S.D. \& U.M. \\
  58991.45731 &  0.00075  &   A  &   S   &   R.U.   &  52509.33511 & 0.00026  &   A  &   P  &   S.D. \& U.M. \\
  59023.47973 &  0.00039  &   B  &   P   &   R.U.   &  52510.40944 & 0.00040  &   A  &   S  &   S.D. \& U.M. \\
  59043.52943 &  0.00062  &   A  &   S   &   R.U.   &  52735.40298 & 0.00091  &   A  &   S  &   S.D. \& U.M. \\
  59040.37413 &  0.00091  &   A  &   P   &   R.U.   &  52825.97834 & 0.00148  &   A  &   P  &   S.D. \& U.M. \\
  59040.52213 &  0.00085  &   B  &   P   &   R.U.   &  52827.05749 & 0.00202  &   A  &   S  &   S.D. \& U.M. \\
  59089.36019 &  0.00045  &   A  &   S   &   R.U.   &  53102.05902 & 0.00052  &   A  &   S  &   S.D. \& U.M. \\
  59102.40213 &  0.00043  &   B  &   S   &   R.U.   &  53171.80904 & 0.00057  &   A  &   P  &   S.D. \& U.M. \\
  59343.51031 &  0.00033  &   A  &   S   &   R.U.   &  52687.49036 & 0.00950  &   A  &   S  &   S.D. \& U.M. \\
  59349.50020 &  0.00025  &   B  &   S   &   R.U.   &  59831.47733 & 0.00027  &   B  &   P  &   FRAM         \\
  59361.46286 &  0.00064  &   B  &   P   &   FRAM   &  59840.35162 & 0.00046  &   A  &   P  &   R.U.         \\
  59367.43420 &  0.00017  &   A  &   P   &   R.U.   &  54669.45443 & 0.00100  &   C  &   P  &   MR13 \\
  59369.51977 &  0.00030  &   A  &   P   &   M.M.   &  59101.29077 & 0.00531  &   C  &   P  &   R.U. \\
  59392.43409 &  0.00019  &   A  &   P   &   R.U.   &  59150.29198 & 0.00134  &   C  &   P  &   R.U. \\
  59416.42568 &  0.00040  &   A  &   S   &   FRAM   &  59152.25463 & 0.00389  &   C  &   P  &   R.U. \\
  59419.52306 &  0.00038  &   A  &   P   &   R.U.   &  59279.67396 & 0.00770  &   C  &   P  &   R.U. \\
  59425.36003 &  0.00070  &   B  &   P   &   R.U.   &  59332.58531 & 0.00473  &   C  &   P  &   R.U. \\
  59463.27078 &  0.00245  &   A  &   P   &   R.U.   &  59338.46323 & 0.00633  &   C  &   P  &   R.U. \\
  59465.35947 &  0.00037  &   A  &   P   &   R.U.   &  59436.47808 & 0.00354  &   C  &   P  &   R.U. \\
  59677.51573 &  0.00055  &   B  &   S   &   R.U.   &  59438.44039 & 0.00257  &   C  &   P  &   R.U. \\
  59679.53400 &  0.00044  &   B  &   P   &   R.U.   &  59497.24210 & 0.01257  &   C  &   P  &   R.U. \\
  59717.43638 &  0.00055  &   A  &   P   &   R.U.   &  59675.60538 & 0.00101  &   C  &   P  &   R.U. \\
  59718.50933 &  0.00034  &   A  &   S   &   R.U.   &  59681.49018 & 0.00107  &   C  &   P  &   R.U. \\
  59742.43700 &  0.00079  &   A  &   P   &   R.U.   &  59779.49300 & 0.00100  &   C  &   P  &   R.U. \\
  59767.43486 &  0.00021  &   A  &   P   &   R.U.   &  52819.10680 & 0.00232  &   C  &   P  &   S.D. \& U.M. \\
  59767.57676 &  0.00069  &   B  &   P   &   R.U.   &  52693.65303 & 0.00273  &   C  &   P  &   S.D. \& U.M. \\
  59775.49098 &  0.00054  &   B  &   S   &   R.U.   &  59828.49557 & 0.00325  &   C  &   P  &   FRAM  \\
  59787.45377 &  0.00079  &   B  &   P   &   R.U.   &  59830.45265 & 0.00304  &   C  &   P  &   FRAM  \\
  59791.42428 &  0.00071  &   A  &   S   &   R.U.   &  \\
 \hline \hline
\label{tbl:ecl-times}
\end{tabular} \\
{\bf Notes:}  G.P. = Gerald Persha, see http://var2.astro.cz ; S.D. \& U.M. = S. Dallaporta \&
U.Munari; R.U. = R.Uhla\v{r}; M.M. = M.Ma\v{s}ek; MR13 = \cite{2013ASSP...31P..39M}
\end{table*}

\end{document}